\documentclass[11pt]{article}
\usepackage{subfigure}
\usepackage{graphicx}
\usepackage{hyperref}
\usepackage{epsfig,subfigure}
\usepackage{color}
\def\QED{\mbox{\rule[0pt]{1.5ex}{1.5ex}}}
\def\proof{\noindent\hspace{2em}{\it Proof: }}

\makeatletter
\def\bbordermatrix#1{\begingroup \m@th
  \@tempdima 4.75\p@
  \setbox\z@\vbox{%
    \def\cr{\crcr\noalign{\kern2\p@\global\let\cr\endline}}%
    \ialign{$##$\hfil\kern2\p@\kern\@tempdima&\thinspace\hfil$##$\hfil
      &&\quad\hfil$##$\hfil\crcr
      \omit\strut\hfil\crcr\noalign{\kern-\baselineskip}%
      #1\crcr\omit\strut\cr}}%
  \setbox\tw@\vbox{\unvcopy\z@\global\setbox\@ne\lastbox}%
  \setbox\tw@\hbox{\unhbox\@ne\unskip\global\setbox\@ne\lastbox}%
  \setbox\tw@\hbox{$\kern\wd\@ne\kern-\@tempdima\left[\kern-\wd\@ne
    \global\setbox\@ne\vbox{\box\@ne\kern2\p@}%
    \vcenter{\kern-\ht\@ne\unvbox\z@\kern-\baselineskip}\,\right]$}%
  \null\;\vbox{\kern\ht\@ne\box\tw@}\endgroup}
\makeatother

\usepackage{arydshln}
\usepackage{amsmath,amsfonts,amssymb}
\usepackage{verbatim}
\usepackage[bookmarks=false]{}
\usepackage[font={small}]{caption}

\newtheorem{theorem}{\bf Theorem}

\newtheorem{definition}{Definition}
\newtheorem{lemma}{Lemma}

\newcommand\blfootnote[1]{%
  \begingroup
  \renewcommand\thefootnote{}\footnote{#1}%
  \addtocounter{footnote}{-1}%
  \endgroup
}

\begin{document}
\date{}

\title{Replication-based Outer bounds and the Optimality of ``Half the Cake'' for Rank-Deficient MIMO Interference Networks}
\author{\normalsize Bofeng Yuan, Hua Sun and Syed A. Jafar\\
\\Center for Pervasive Communications and Computing (CPCC)\\
University of California, Irvine\\
Email: $\left\{\right.$bofengy, huas2, syed$\left.\right\}$@uci.edu}

%\author{\normalsize Bofeng Yuan, Hua Sun and Syed A. Jafar}

\maketitle
\blfootnote{Presented in part at IEEE GLOBECOM 2015 \cite{BofengGC15}.}
%\blfootnote{Presented in part at IEEE GLOBECOM 2015 \cite{BofengGC15}. This work was supported in part by ONR grant N00014-15-1-2557, and by NSF grants CCF-1319104 and CCF-1161418.}

\begin{abstract}
In order to gain new insights into MIMO interference networks, the optimality of $\sum_{k=1}^K M_k/2$ (half the cake per user) degrees of freedom is explored for a $K$-user multiple-input-multiple-output (MIMO) interference channel where the cross-channels have arbitrary rank constraints, and the  $k^{th}$ transmitter and receiver are equipped with $M_k$ antennas each. The result consolidates and significantly generalizes results from prior studies by Krishnamurthy et al., of rank-deficient interference channels where all users have $M$ antennas;  and by Tang et al., of full rank interference channels where the $k^{th}$ user pair has $M_k$ antennas. The broader outcome of this work is a novel class of replication-based outer bounds for arbitrary rank-constrained MIMO interference networks where replicas of existing users are added as auxiliary users and the network connectivity is chosen to ensure that any achievable scheme for the original  network also works in the new network. The replicated network creates a new perspective of the problem, so that even simple arguments such as user cooperation become quite powerful when applied in the replicated network, giving rise to stronger outer bounds, than when applied directly in the original network. Remarkably, the replication based bounds are broadly applicable not only  to  MIMO interference channels with arbitrary rank-constraints, but much more broadly, even beyond Gaussian settings.
\end{abstract}

\section{Introduction}
\begin{figure}[h]
%\begin{center}
\centering 
\includegraphics[width=3.5in]{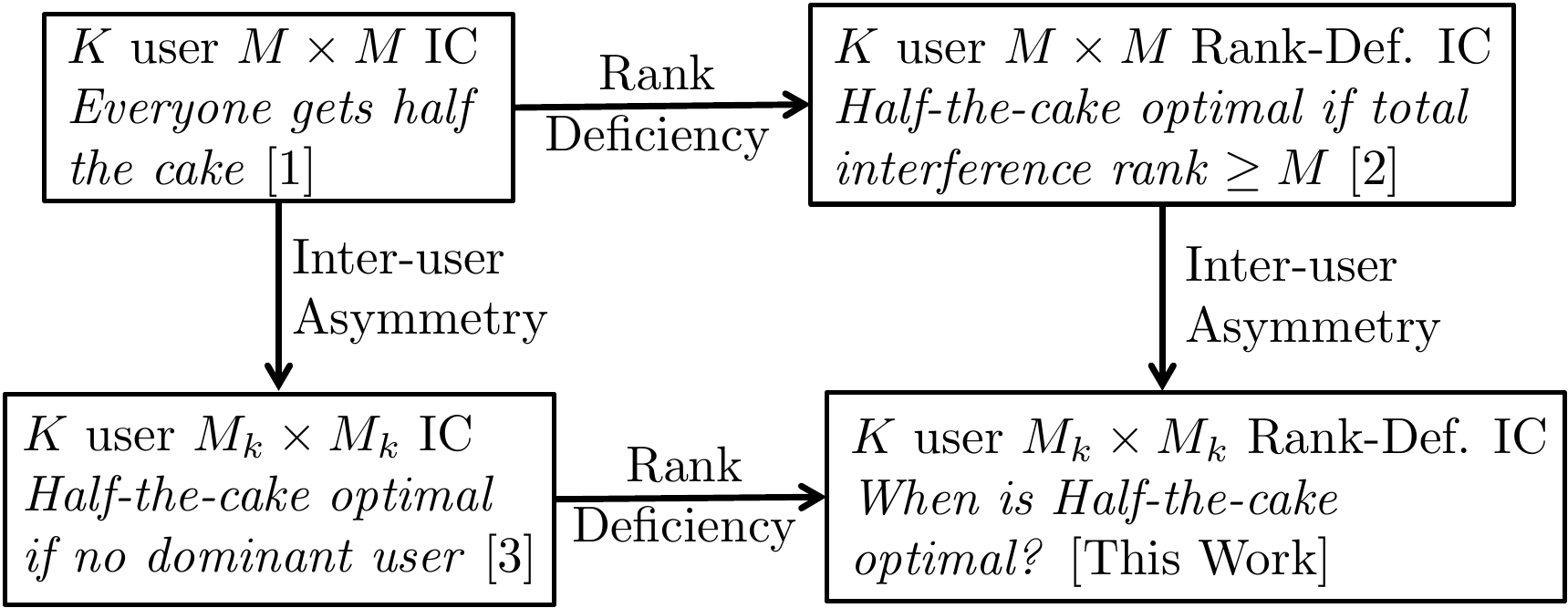}
\caption{The motivation of this paper. IC stands for interference channel. Rank-Deficient is abbreviated as Rank-Def. .}\label{motivation}
%\end{center}
\end{figure}

Degrees of freedom (DoF) studies of wireless interference networks have produced a diverse array of new insights into the accessibility of signal dimensions under a variety of channel models. In order to consolidate these insights and to build upon them, it is important to make progress on unifying the underlying channel models. The motivation for this work, summarized in Fig. \ref{motivation},  is to pursue such a generalization of the results from \cite{cadambe2008interference, krishnamurthy2015degrees,liu2014dof}. Specifically, in this work we start with the goal of consolidating the key insights regarding the optimality of half-the-cake (the ``cake" refers to each user's interference-free DoF, cf. \cite{cadambe2008interference}) for the $K$-user MIMO interference channel settings where the number of antennas at each receiver is equal to the number of antennas at the corresponding transmitter, i.e., all the \emph{desired} channels are \emph{square} matrices. %\footnote{Rectangular channels present a significantly different set of challenges, and generally allow more than half-the-cake per user, so they remain outside the scope of this paper. (? Seems we indeed consider rectangular channels, although not for half-the-cake optimality problem. Should we delete this footnote?)} 
The study of the unified setting leads us to a broader outcome -- a novel class of replication-based outer bounds that are applicable not only to arbitrary rank-constrained MIMO interference networks but much more generally, even beyond Gaussian settings as well. 

\subsection{Everyone Gets Half the Cake}
 It was shown by Cadambe and Jafar in \cite{cadambe2008interference} that in a $K$-user $M\times M$ MIMO interference channel where each node is equipped with $M$ antennas,  the optimal DoF value is $KM/2$. Since each user achieves half of his interference-free DoF, the result is often paraphrased as ``everyone gets half the cake". Generalizations of this result have been explored in various directions, in particular to find out when the optimal solution may allow even more than half-the-cake. Indeed rectangular interference channels (cf.  \cite{gou2010degrees, ghasemi2011interference, wang2014subspace, bresler2014interference, wang2014genie}), and multi-hop settings (cf. \cite{gou2012aligned}) have shown that more than half-the-cake is possible. Of particular interest to us in this work are the generalizations in \cite{krishnamurthy2015degrees,liu2014dof}.
 
\subsection{Optimality of Half-the-cake: Key Insight from \cite{krishnamurthy2015degrees, liu2014dof}}
The generalization in \cite{krishnamurthy2015degrees} concerns rank deficient channels. Rank deficient interference channels (cf. \cite{chae2011degrees, chae2013degrees, zeng2014degrees}) are frequently encountered due to poor scattering, keyhole effects, as well as underlying topological and structural concerns in single-hop abstractions of multihop networks with linear forwarding at intermediate nodes. Cross-channel rank-deficiencies have the potential to be helpful as the scope of zero forcing schemes is enhanced (although the scope of interference alignment schemes is limited by rank-deficiencies), opening the possibility that more than half-the-cake may be achievable. Exploring this possibility in  \cite{krishnamurthy2015degrees}, Krishnamurthy and Jafar establish that for the $K$-user $M\times M$ MIMO interference channel where all the cross channels are rank-deficient with the same rank $D\leq M$ and direct channels are full rank, $KM/2$ DoF (half-the-cake) are  optimal if the sum of all interference ranks at each user, is greater than or equal to the number of antennas at the user, $(K-1)D\geq M$. In other words, \emph{every signal dimension is accessible by at least one interfering user}. For $K=3$ users, \cite{krishnamurthy2015degrees} considers a more general setting, so that at each receiver the interfering channel from the preceding transmitter is of rank $D_1$ and the interfering channel from the next transmitter (with wrap around) is $D_2$. For $K=2$ users the setting is fully general with all interfering channel ranks allowed to take arbitrary values. Remarkably, in all cases, the key insight remains the same: 

{\bf Original Insight:} ``\emph{Half-the-cake is optimal if at every transmitter and receiver, the sum of interfering channel ranks is greater than or equal to the number of antennas at that transmitter and receiver, respectively.}"

Finally, Liu, Tuninetti and Jafar in \cite{liu2014dof} consider a different generalization, to the $K$-user $M_k\times M_k$ MIMO interference channel with full rank generic channels, where the $k^{th}$ user has $M_k$ transmit and $M_k$ receive antennas. For this setting \cite{liu2014dof} showed that half-the-cake is optimal provided there is no  dominant user (a user with more antennas than all the rest of the users combined). Interestingly,  this condition is also identical to the insight from \cite{krishnamurthy2015degrees} --- once again, half-the-cake is optimal if the sum of interfering channel ranks  is greater than or equal to the number of antennas at each user.

\subsection{Overview}\label{overview}
In order to further refine the key insight from \cite{krishnamurthy2015degrees,liu2014dof} and to identify its limitations, it is important to continue to test its validity under generalized settings. To this end, in this work we unify the channel models of \cite{krishnamurthy2015degrees} and \cite{liu2014dof} into the rank-deficient $K$-user $M_k\times M_k$ MIMO interference channel, and study the optimality of half-the-cake under \emph{arbitrary} (no assumptions of symmetry) rank constraints on the cross-channels. 

Surprisingly, we discover that the original insight fails in this generalized setting. Indeed, as a \emph{counterexample} consider the $3$-user MIMO interference channel with $M_1=10, M_2=8, M_3=6$, where the channel from Transmitter 1 to Receiver 2 has rank $5$ and the channel from Transmitter 2 to Receiver 1 has rank $6$. All other channels have full rank. Even though  in this channel, the sum of interfering channel ranks at every user is greater than or equal to the number of antennas at that user, it is possible to achieve more than half-the-cake  (half-the-cake is 12, but 12.5 DoF are achievable, as explained in Appendix \ref{sec:example}). Therefore, a new outer bound is  necessary for the $K$-user $M_k\times M_k$ MIMO interference channel. 

Define $M_\Sigma = \sum_{k=1}^K M_k$. %$M_\Sigma=\Sigma_{k\in\mathcal{I}_K}M_k$. 
Define ${\bf H}_{ji}$ as the $M_j \times M_i$ channel matrix from Transmitter $i$ to Receiver $j$, $i, j \in \{1,\cdots, K\}$.
Define ${\bf H}$ as the overall $M_\Sigma \times M_\Sigma$ channel matrix from all  $K$ transmitters to all $K$ receivers (i.e., $([{\bf H}_{ji}])$), and let $\bar{\bf H}$ be obtained from ${\bf H}$ by replacing all desired channels (i.e., channels between corresponding transmitter-receiver pairs, ${\bf H}_{kk}$) with zeros. 
Our new insight for the unified setting comes from a novel outer bound argument that shows that the DoF cannot exceed half-the-cake if $\bar{\bf H}$ has full rank. 
In light of our outer bound, the counterexample mentioned above implies that the $24\times 24$ matrix 
\begin{eqnarray*}
{\bf \bar{H}}=\bordermatrix{&\mbox{\tiny 10}&\mbox{\tiny 8}&\mbox{\tiny 6}\cr
\mbox{\tiny 10}&{\bf 0}&{\bf H}_{12}&{\bf H}_{13}\cr
 \mbox{\tiny 8}&{\bf H}_{21}&{\bf 0}&{\bf H}_{23}\cr
 \mbox{\tiny 6}&{\bf H}_{31}&{\bf H}_{32}&{\bf 0}\cr
 },
 \mbox{ with ranks }
 \bordermatrix{&\mbox{\tiny 10}&\mbox{\tiny 8}&\mbox{\tiny 6}\cr
\mbox{\tiny 10}&0&6&6\cr
 \mbox{\tiny 8}&5&0&6\cr
 \mbox{\tiny 6}&6&6&0\cr
 }
\end{eqnarray*}
cannot have full rank for any possible realization. Indeed, this is the case because the $24\times 18$ sub-matrix formed by its first $18$ columns is rank-deficient (sum of row ranks cannot be more than $6+5+6=17$).

Stated in an equivalent form, the new outer bound leads us to a more precise understanding of the original insight, so that we are able to refine it to the following form for generic rank-deficient channels.

{\bf Refined Insight:} ``\emph{Half-the-cake is optimal if at every transmitter and receiver,  the sum of {\bf  reduced} interfering channel ranks {\bf equals} the number of antennas at that  transmitter and receiver, respectively.}"

So according to the refined condition, we are allowed to reduce the ranks of the cross-channels, but the reduced interference channel ranks must then add up at each transmitter and receiver to precisely equal the number of antennas at that transmitter and receiver, respectively. The counterexample presented earlier does not satisfy the refined condition. Indeed, it is not possible to assign any (possibly reduced) rank values that add up to the row and column index for every row and every column. %A precise statement of the main result appears in Theorem \ref{theorem:main} in Section \ref{sec_result} of this paper.

On the other hand, consider a different $\bar{\bf H}$ with ranks
\begin{eqnarray*}
 \bordermatrix{&\mbox{\tiny 10}&\mbox{\tiny 8}&\mbox{\tiny 6}\cr
\mbox{\tiny 10}&0&8&3\cr
 \mbox{\tiny 8}&5&0&4\cr
 \mbox{\tiny 6}&6&2&0\cr
 }\mbox{ which can be reduced to }
  \bordermatrix{&\mbox{\tiny 10}&\mbox{\tiny 8}&\mbox{\tiny 6}\cr
\mbox{\tiny 10}&0&8&2\cr
 \mbox{\tiny 8}&4&0&4\cr
 \mbox{\tiny 6}&6&0&0\cr
 }
\end{eqnarray*}
so that the reduced ranks add up to the row and column index for every row and column. Therefore,  any realization of $\bar{\bf H}$ channels with these (unreduced or reduced) ranks cannot achieve more than half-the-cake. Also, as we show, for generic channels half-the-cake is always achievable, so it is optimal.

As a ``sufficient" condition for optimality of half-the-cake, the additional requirements in the refined condition may appear to weaken its impact. This is not the case, however, as we note that the refined condition still recovers all prior results on the optimality of half-the-cake from \cite{cadambe2008interference, krishnamurthy2015degrees,liu2014dof} as special cases of the $K$-user $M_k\times M_k$ rank-deficient MIMO channel model. For $K=3$, we also show that if the rank of each interference link is symmetric, i.e., $\mbox{rank}\,({\bf H}_{ji})=\mbox{rank}\,({\bf H}_{ij})$, then the condition is also necessary for half-the-cake DoF to be optimal. 

The broader technical contribution of this work is a novel class of replication-based DoF outer bounds that are applicable to the general  $K$-user $M_k\times N_k$ MIMO interference channel with arbitrary rank-constraints, where all the nodes can have different number of antennas. The DoF  of general MIMO interference channels are of fundamental interest as they shed light into the accessibility of signal dimensions with local joint processing (MIMO) at each node within the globally distributed setting that is an interference network. In particular, information theoretic DoF outer bounds for MIMO interference channels offer a powerful tool beyond  the cut-set bounds used extensively in the study of wireless and wired communication networks. As such, DoF outer bounds  have been studied in \cite{jafar2007degrees, gou2010degrees, wang2014subspace, torrellas2014dof, liu2014dof, wang2014genie, liu2013genie}, mostly for symmetric settings, leading to various approaches based on cooperation \cite{gou2010degrees}, change of basis operations \cite{wang2014subspace, torrellas2014dof, liu2014dof} and  genie-chains  \cite{wang2014genie, liu2013genie}. However, in spite of much progress, the DoF of MIMO interference networks remain unknown in general, even in symmetric settings, but especially under asymmetric settings. Evidently, there is a need for new outer bounding arguments to extend, complement, and where possible, simplify the existing approaches. It is in these regards that the new DoF outer bounds developed in this work are significant.

The key step in our replication-based outer bounding approach is to include auxiliary users as copies of existing users with corresponding independent auxiliary messages,  ensure the connectivity is such that any achievable scheme for the original $K$-user network continues to work in the
new network, creating a new network where simple bounds  (such as Carleial's bound in \cite{carleial1983outer} and cooperation based bounds) can be applied to produce various weighted sum-rate bounds for the original network. This approach  provides us a class of outer bounds for general $K$-user $M_k\times N_k$ MIMO interference channel with any given channel realization. While the new bounds are conceptually quite simple and easily extendable to weighted sum-rates, a challenging aspect of these information theoretic bounds is that there could be many valid connectivity patterns that produce distinct outer bounds so that  finding the best bound may be computationally cumbersome. However, this aspect can be greatly simplified if the bounds are restricted to linear DoF, i.e., DoF achieved by \emph{linear} precoding schemes. Remarkably, unlike prior works on feasibility of linear schemes \cite{Yetis_Gou_Jafar_Kayran_TSP, bresler2014interference, ruan2013feasibility, gonzalez2014feasibility} which do not allow symbol extensions or asymmetric signaling and focus on generic channels, the resulting linear DoF outer bounds from our work allow all possible linear schemes (including symbol extensions, asymmetric signaling) and apply to arbitrary interfering channels (not only generic ones).

Finally, we note that while, to the best of our knowledge, this is the first application of replication-based outer bounds to general MIMO interference networks, our ideas for  replication-based outer bounds on weighted sum rates originated from our earlier studies of optimality of treating interference as noise in parallel interference networks \cite{Sun_Jafar_ParallelTIN}. It is also worthwhile to note that during the course of preparing this full paper from the original conference version of this work {\color{black}\cite{BofengGC15}}, we have  come across another very recent independent work in \cite{Kiamari_Avestimehr} which also relies on replication-based outer bounds, in the context of 2 and 3 user symmetric deterministic interference channels. We believe that the convergence toward replication-based bounds from different perspectives points to their fundamental significance. As such, exploring the full potential of replication-based bounds is a promising direction for future work.

The paper is organized as follows. In Section \ref{sec_model}, the system model is introduced. The main results are formally stated in Section \ref{sec_result}. Section \ref{recover} shows that prior results on optimality of half-the-cake can be recovered as special cases of our generalized result (a counterexample to the original insight we mentioned above is presented in Appendix \ref{sec:example}). In Section \ref{sec:two_example},  examples of applications of the new bound are presented. 

{\it Notation:} We denote the set $\left\{1,...,K\right\}$ by $\mathcal{I}_{K}$ for a positive integer $K$. For a subset $A$ of $\mathcal{I}_K$,  $\mathcal{I}_{K}\backslash A$ denotes the set of elements that are in $\mathcal{I}_K$ but not in $A$, e.g., if $A = \{l\}, l \in \mathcal{I}_K$, then  $\mathcal{I}_{K}\backslash\l=\left\{1,...,l-1,l+1,...,K\right\}$. Indexing is interpreted in a circular wrap-around manner, modulo the number of users, e.g., the $K^{th}$ user is same as the $0^{th}$ user. ${\bf I}_{m}$ denotes the $m \times m$ identity matrix and ${\bf 0}_{m_1\times m_2}$ denotes the $m_1 \times m_2$ matrix of zeros.

%\section{$K$-user  Rank-Deficient $(M_k\times M_k)$ Channel}
\section{System Model}\label{sec_model}
\subsection{$K$-user Rank-Constrained MIMO Interference Channel}
The general setting of interest is the $K$-user MIMO interference channel where there are $M_k$ and $N_k$ antennas at the $k^{th}$ transmitter and receiver, respectively. Each transmitter sends an independent message to its corresponding receiver. We refer to this general setting as the $(M_k\times N_k)$ channel. At time slot $t\in\mathbb{Z}^+$, the received signal vector at Receiver $j$ is given by 
\begin{equation}
Y_{j}(t)=\sum_{i=1}^K \mathbf{H}_{ji}(t)X_{i}(t)+Z_{j}(t) 
\end{equation}
where $X_{i}(t)\in\mathbb{C}^{M_{i}\times1}$ is the signal vector sent from Transmitter $i$ which satisfies an average power constraint $\mathbb{E}(\lVert X_{i}(t)\rVert^{2}){\le}\rho$, $Z_{j}(t)\in\mathbb{C}^{N_{j}\times1}$ is the i.i.d. circularly symmetric complex additive white Gaussian noise (AWGN) at Receiver $j$, each entry of which is an i.i.d. Gaussian random variable with zero-mean and unit-variance, and $\mathbf{H}_{ji}(t)\in\mathbb{C}^{N_{j}\times M_{i}}$ is the channel matrix from Transmitter $i$ to Receiver $j$. We assume that perfect global channel knowledge is available at all nodes. 

The desired channel matrices $\mathbf{H}_{ii}(t)$ are assumed to be full rank\footnote{Similar to \cite{krishnamurthy2015degrees}, the extension to  rank-deficient desired channels is straightforward.}  while the cross channels ${\bf H}_{ji}$ are subject to rank constraint $D_{ji}$. By default the channels are assumed to be \emph{generic} --- by which we mean the channels are ergodically time-varying and drawn from continuous distributions subject to rank-constraints. Similar to \cite{krishnamurthy2015degrees}, a rank-constrained generic $N_j\times M_i$ channel matrix of rank $D_{ji}$ is modeled as a product of an $N_j\times D_{ji}$ matrix with a $D_{ji}\times{M_i}$ matrix, all of whose entries are drawn from a continuous distribution. 

We note that our  DoF outer bounds, which are the primary focus of this work, also hold for \emph{arbitrary} channels, i.e., without the assumptions of generic and ergodically time-varying channels. Achievability results are included to highlight the quality of the bounds, which are shown to be tight for generic channels. While we expect the results to hold true (almost surely) even without ergodicity or time-variations, choosing ergodic time-varying channels allows us to simplify the achievability arguments as much as possible, so that the focus of this work remains on the outer bounds.

The achievable rates, capacity region and DoF region of this network are defined in the standard sense (see \cite{cadambe2008interference}). We define the sum-DoF value as $d_{\Sigma}=\lim_{\rho \to \infty}R_{\Sigma}(\rho)/\log(\rho)$, where $R_{\Sigma}(\rho)$ is the maximum sum rate at Signal-to-noise ratio, $\rho$. We also define $N_\Sigma=\Sigma_{k\in\mathcal{I}_K}N_k$, $M_\Sigma=\Sigma_{k\in\mathcal{I}_K}M_k$.

\section{Results}\label{sec_result}
In this section we state the main results of this work. 

\subsection{The Rank-Constrained $K$-user $(M_k\times M_k)$ Interference Channel -- Optimality of Half-the-Cake}
In this section, we focus on the $(M_k\times M_k)$ setting, i.e., where $N_k=M_k$, so that the desired channel matrices are generic full rank square matrices, while the interference channel matrices are in general rectangular and  subject to arbitrary rank-constraints. This setting unifies and generalizes the cases studied in \cite{krishnamurthy2015degrees} and \cite{liu2014dof}, and forms our starting point. We start with the achievability result, which is a simple application of the ideas of ergodic interference alignment \cite{nazer2012ergodic} and blind interference alignment \cite{Jafar_BlindIA}, which says in this case, that  for generic channels, ``half-the-cake" is almost surely achievable.

\begin{theorem}\label{theorem:ach}
{\it For generic channels,  regardless of interference rank-constraints
\begin{eqnarray*}
%\mbox{DoF}
d_{\Sigma} \geq M_\Sigma/2
\end{eqnarray*}
}
\end{theorem}

\proof Since the channels are ergodically time-varying and drawn from continuous distributions, %in the long run 
we may partition the channels over all time slots to pairs of 2 channel uses, such that for each 2 channel uses, say at times $t_1$ and $t_2$,  all channel matrices of interference links remain the same $\mathbf{H}_{ji}(t_1)=\mathbf{H}_{ji}(t_2)$, $i\ne j$, and all channel matrices of direct links change $\mathbf{H}_{ii}(t_1)\ne\mathbf{H}_{ii}(t_2)$ in a generic sense, i.e., their difference is also full rank. Then each transmission takes such 2 channel uses, and by letting each transmitter repeat its symbols over the 2 channel uses, each receiver can eliminate interference by subtracting the output at $t_2$ from the output at $t_1$, and obtain an $M_k\times M_k$ interference free channel, over which $M_k$ DoF are obtained. Since, this requires two channel uses, effectively $\tfrac{M_k}{2}$ DoF are achieved for User $k$ and in total $M_\Sigma/2$ sum-DoF are achieved. 
\hfill \QED

The main question of interest is, when is half-the-cake \emph{optimal}? To answer this we introduce a new replication-based outer bound argument that will turn out to be quite broadly applicable. Recall that ${\bar {\bf H}}$ is the overall $M_\Sigma\times M_\Sigma$ channel matrix where all desired channels ${\bf H}_{kk}$ have been set to zero. Specialized to our present purpose, the outer bound is presented below.
\begin{theorem}\label{theorem:LDoF}
{\it For arbitrary channel realizations, if 
\begin{eqnarray*}
\mbox{rank}(\bar{\bf H})=M_\Sigma& \mbox{ then } & d_\Sigma \leq M_\Sigma/2.
\end{eqnarray*}}
\end{theorem}
Note that the outer bound applies to arbitrary channels, i.e., without any requirements for time-varying, ergodic, or generic realizations. Remarkably, the proof is quite simple, based upon a replication argument.

\begin{figure}[h] 
\centering
\subfigure[]{
%\label{3user1}
\includegraphics[width=0.34\textwidth]{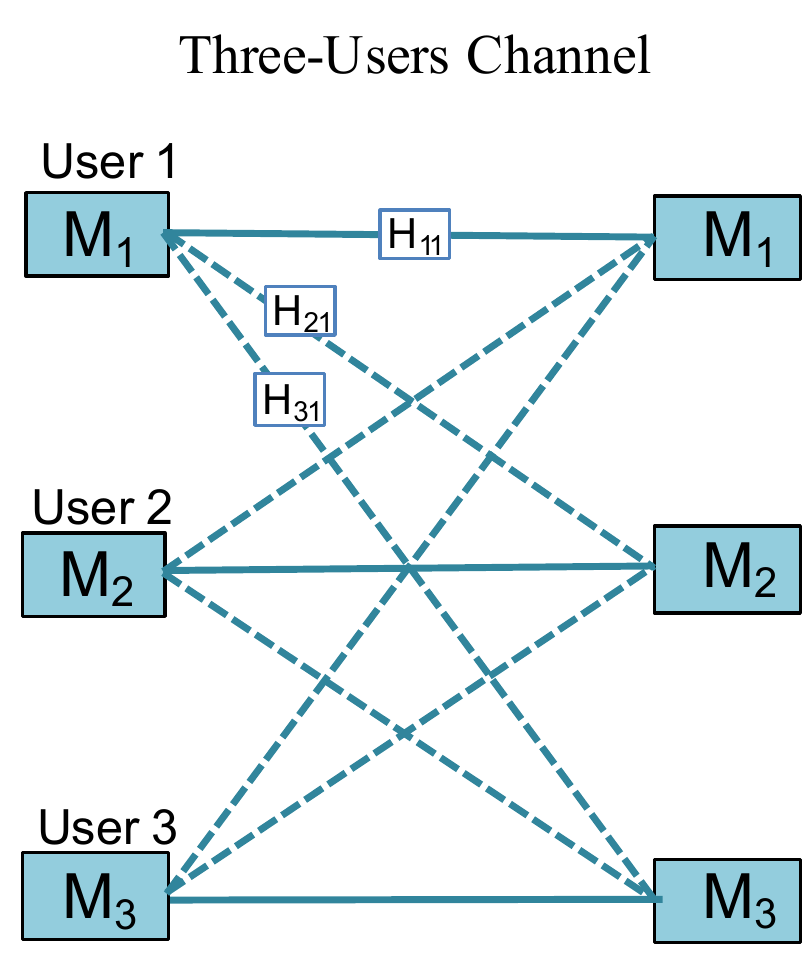}}
\hspace{1.2in}
\subfigure[]{
%\label{3user2}
\includegraphics[width=0.35\textwidth]{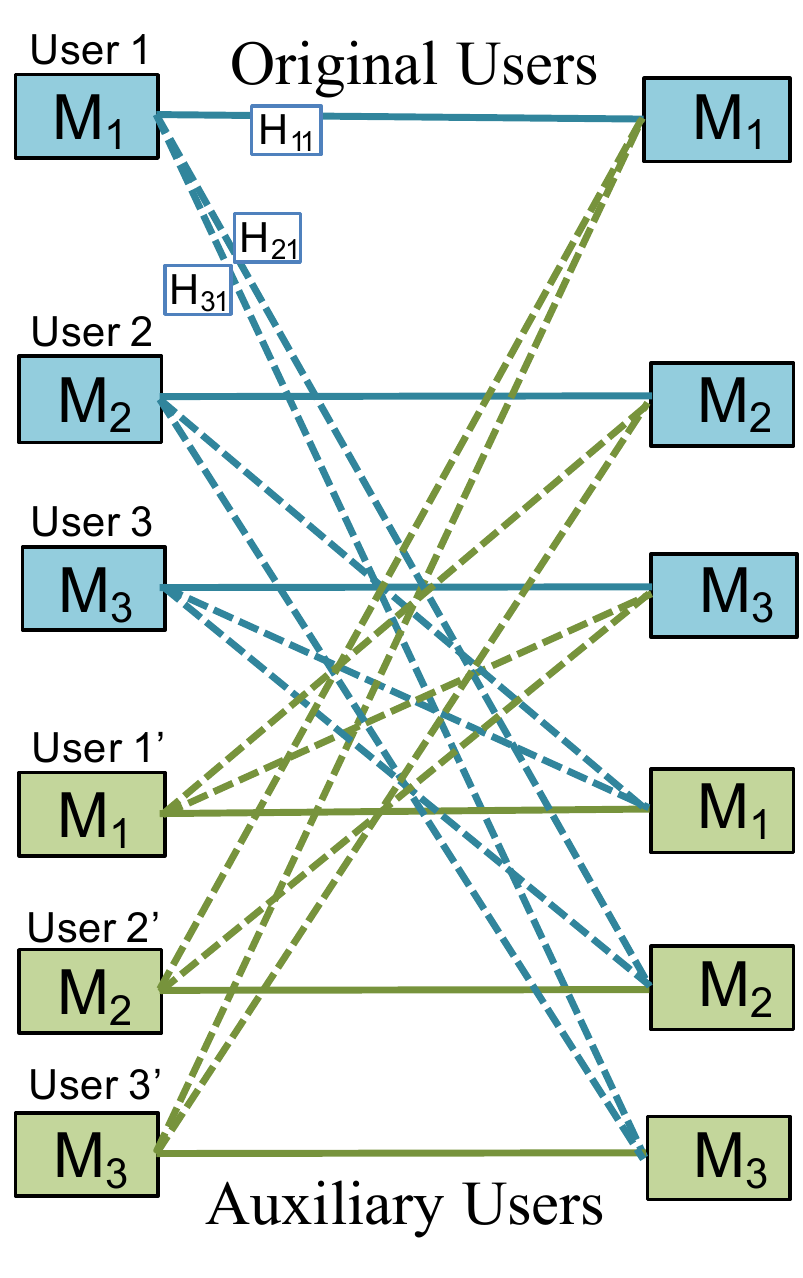}}
\caption[]{
\subref{3user1} A 3-user interference channel, and \subref{3user2} The $6$-user interference channel created in Theorem \ref{theorem:LDoF}.}
\label{3case}
\end{figure}

\proof Given the original $K$-user interference channel with channel matrices ${\bf H}_{ji}$, now create a $2K$-user interference channel by adding an auxiliary User $k'$ for each Original User $k$. We denote the channels in the new $2K$-user network by notations with hat symbol,
%In order to better distinguish with the channel matrix ${\bf H}_{ij}$, $ \forall i,j\in\mathcal{I}_K$ in the original channel, let us use $\hat{\bf H}$ to denote each channel matrix in this new $2K$-user network, 
e.g., $\hat{\bf H}_{ji'}$ represents the channel matrix from Transmitter $i'$ to Receiver $j$. The new channels are chosen so that $ \forall i,j \in\mathcal{I}_K$,  1) $\hat{\bf H}_{j'i}=\hat{\bf H}_{ji'}= {\bf H}_{ji}$ whenever $i\neq j$, 2) $ \hat{\bf H}_{i'i'}=\hat {\bf H}_{ii}= {\bf H}_{ii}$, 3) $ \hat{\bf H}_{j'i'}=\hat{\bf H}_{ji}$ is the matrix of zeros whenever $i\neq j$, and 4) $\hat{\bf H}_{i'i}=\hat{\bf H}_{ii'}$ is the matrix of zeros. For a pictorial illustration of the case where $K=3$, see Fig. \ref{3case}.

Any coding scheme for the original channel still works if each auxiliary User $i'$ uses the same codebook as User $i$. 
%The reason is that the received signal at each user in the new network is statistically the same as the corresponding user in the original network. 
Since Users $i$ and $i'$ in the new network achieve the same rates as User $i$ in the original network, the sum-DoF value for the new network is at least twice that of the original network. Now in the new network, allow all original transmitters to cooperate, all original receivers to cooperate, all auxiliary transmitters to cooperate and all auxiliary receivers to cooperate, which can only help. This creates a 2-user interference channel where everyone has $M_\Sigma$ antennas, and where the interference matrix is $\bar{\bf H}$. If this interference matrix is full rank, then each user, after decoding its desired signal, can subtract it out and then proceed to decode the interfering signal as well (subject to noise distortion, inconsequential for DoF). Thus, the sum-DoF of the interference channel cannot be more than $M_\Sigma$, and therefore the sum-DoF of the original network cannot be more than $\frac{1}{2}M_\Sigma$.\hfill\QED

For generic channels, the condition of Theorem \ref{theorem:LDoF} can be presented in a simpler alternative form, in terms of the ranks of the  individual interfering channels, as follows.
\begin{lemma}\label{lemma:connect}
{\it For generic channel realizations, $\mbox{rank}(\bar{\bf H})=M_\Sigma$ if and only if there exist reduced ranks $\bar{D}_{ji}\leq D_{ji}$ for each interference link, which satisfy the following condition, 
\begin{equation}\label{lemma_eq}
\sum_{j\in\mathcal{I}_{K}\backslash i}\bar{D}_{ji}=\sum_{j\in\mathcal{I}_{K}\backslash i}\bar{D}_{ij}=M_i, \forall i \in\mathcal{I}_{K}.
\end{equation}
}
\end{lemma}
The proof of Lemma \ref{lemma:connect} is presented in Appendix \ref{proof:connect}.

Combined with Theorem \ref{theorem:LDoF}, Lemma \ref{lemma:connect} directly proves the following theorem, which unifies and generalizes the results from \cite{krishnamurthy2015degrees} and \cite{liu2014dof}.
\begin{theorem}\label{theorem:main}
{\it For a $K$-user generic rank-deficient MIMO interference channel, if there exist reduced ranks $\bar{D}_{ji}\leq D_{ji}$ for each interference link, which satisfy the following condition, 
\begin{equation}\label{1}
\sum_{j\in\mathcal{I}_{K}\backslash i}\bar{D}_{ji}=\sum_{j\in\mathcal{I}_{K}\backslash i}\bar{D}_{ij}=M_i, \forall i \in\mathcal{I}_{K}.
\end{equation}
then almost surely half-the-cake is optimal, i.e., 
$%\mbox{DoF}
d_\Sigma =\sum_{k=1}^K \frac{M_k}{2}.$
%and for $K=3$, 
%\begin{equation}
%\mbox{DoF}=\sum_{k=1}^K \frac{M_k}{2}.
%\end{equation}
}
\end{theorem}

{\it Remark:} For 3 users, one can state Condition (\ref{1})  more explicitly as follows.
\begin{equation}\label{5}
\min\left\{M_1+D_{32}, M_2+D_{13}, M_3+D_{21}\right\}+\min\left\{M_3+D_{12}, M_1+D_{23}, M_2+D_{31}\right\}\ge M_1+M_2+M_3.
\end{equation}

Theorem \ref{theorem:main} presents a sufficient condition for the optimality of half-the-cake in generic settings. The condition is not a necessary condition for the optimality of half-the-cake. However, combined with the achievability result of Theorem \ref{theorem:ach}, it  recovers the corresponding results from \cite{krishnamurthy2015degrees} and \cite{liu2014dof}. Finding a  condition that is both necessary and sufficient seems to be a difficult task in general, mainly due to the abundance of distinct parameter regimes. The following theorems offer interesting insights into this.

\begin{theorem}\label{theorem:new}
{\it For a 3-user generic rank-deficient MIMO interference channel, if the rank of each interference link is symmetric, i.e., $D_{ji}=D_{ij}$, then the condition in Theorem \ref{theorem:main} is necessary and sufficient for half-the-cake  to be optimal.
%and for $K=3$, 
%\begin{equation}
%\mbox{DoF}=\sum_{k=1}^K \frac{M_k}{2}.
%\end{equation}
}
\end{theorem}
The proof of Theorem \ref{theorem:new} appears in Appendix \ref{proof:new}.

The following two theorems show that  Condition \eqref{1} is not necessary for the optimality of half-the-cake. The proofs are presented in Appendix \ref{proof:last}.

\begin{theorem}\label{special2}
For a 3-user generic rank-deficient MIMO interference channel where $M_1=M_2+M_3$, half-the-cake is optimal, i.e., the sum-DoF value is $\frac{1}{2}M_\Sigma$ if the following condition is satisfied
\begin{eqnarray}\label{21}
D_{12} = M_2, D_{13} = M_3 ~~\mbox{or} ~~D_{21} = M_2, D_{31} = M_3
%\D_{1i}=M_i,\quad\mbox{or}\quad D_{i1}=M_i,\quad\forall i \in\left\{2, 3\right\}. 
%\\
%\mbox{Do we mean this?} (D_{1i}=M_i, \quad\forall i \in\left\{2, 3\right\}) \quad\mbox{or} \quad (D_{i1}=M_i,\quad\forall i \in\left\{2, 3\right\})
\end{eqnarray}
\end{theorem}

{\it Remark: }To see how the condition in Theorem \ref{special2} violates Condition \eqref{1}, consider the example where $M_1=5$, $M_2=3$ and $M_3=2$, (i.e., $M_1 = M_2 + M_3$), $D_{21} = M_2 = 3$, $D_{31} = M_3 = 2$, and all other interference channel matrices are matrices of zeros. Note that this example satisfies the condition in Theorem \ref{special2}.
%As it will show in Section \ref{subsec_specialcase2}, half-the-cake is optimal for this channel. 
However, since $D_{12}=D_{13}=0$, there are no reduced ranks $\bar{D}_{12}$ and $\bar{D}_{13}$ such that $\bar{D}_{12}+\bar{D}_{13}=M_1=5$, i.e., Condition (\ref{1}) is violated.

\begin{theorem}\label{general3user}
For a 3-user generic rank-deficient MIMO interference channel where $M_1=M_2$, half-the-cake is optimal, i.e., the sum-DoF value is $\frac{1}{2}M_\Sigma$, if the following condition is satisfied
\begin{equation}\label{eq:3user_case2}
%\begin{split}
D_{21}=M_{1}, D_{31}=D_{23}=M_{3} ~~or~~ %\\
D_{12}=M_{1}, D_{13}=D_{32}=M_{3}
%\end{split}
\end{equation}
\end{theorem}

{\it Remark:} To see how the condition in Theorem \ref{general3user} violates Condition \eqref{1}, consider the example where $M_1=M_2=5$ and $M_3=3$, $D_{21}= M_1 = 5$ and $D_{31}=D_{23} = M_3 = 3$, and all other interference channel matrices are matrices of zeros. Note that this example satisfies the condition in Theorem \ref{general3user}. However, since $D_{12} = D_{13}=0$, there are no reduced ranks $\bar{D}_{12}$ and $\bar{D}_{13}$ such that $\bar{D}_{12}+\bar{D}_{13}=M_1=5$, i.e., Condition (\ref{1}) is violated. 

\subsection{Replication-Based Bounds for General $(M_k\times N_k)$  Rank-Constrained $K$-user Interference Channel}
As discussed previously, the outer bound that we introduce in Theorem \ref{theorem:LDoF}, is of particular interest in and of itself as it is based on a rather broadly applicable replication argument. The simplicity of this argument makes it easy to generalize the outer bounds. To emphasize this point, in this section we consider some generalizations of the outer bound to the $(M_k\times N_k)$ channel. For this, we first define a new $(\mu_1 + \mu_2 + \cdots + \mu_K)$-user ``replicated" network as follows.

\begin{definition}\label{def::new_network} {\bf [Replicated Network]}
For any given $(M_k\times N_k)$ channel described by channel matrices ${\bf H}_{ji}$,  we create a new $(\mu_1 + \mu_2 + \cdots + \mu_K)$-user interference channel by replacing each User $i$ with $\mu_i$ auxiliary users (replicas), and denoting them as User $i^{[1]}$, User $i^{[2]}$, $\cdots$, User $i^{[\mu_i]}$, each with its own independent message. In this replicated network, we denote the channel matrix from Transmitter $i^{[\alpha]}$ to Receiver $j^{[\beta]}$ as $\hat{\bf H}_{j^{[\beta]}i^{[\alpha]}}$, $ \forall i,j\in\mathcal{I}_K, \alpha \in\mathcal{I}_{\mu_i}, \beta \in \mathcal{I}_{\mu_j}$.
The channel matrices in the replicated network are chosen to satisfy the following constraints. 

1) $\forall i, \alpha, \hat{\bf H}_{i^{[\alpha]}i^{[\alpha]}}= {\bf H}_{ii}$ %and $\forall \gamma \in \mathcal{I}_{\mu_i} \neq \alpha$, $\hat{\bf H}_{i^{[\alpha]}i^{[\beta]}}$ is the matrix of zeros, 
and $\forall \gamma \in \mathcal{I}_{\mu_i}, \gamma \neq \alpha, \hat{\bf H}_{i^{[\gamma]}i^{[\alpha]}}$ are matrices of zeros,

2) $\forall i \neq j$, $\forall \beta$, there exists an $\alpha$ such that $\hat{\bf H}_{j^{[\beta]}i^{[\alpha]}}={\bf H}_{ji}$ and $\forall \gamma \in \mathcal{I}_{\mu_i}$, $\gamma \neq \alpha$, $\hat{\bf H}_{j^{[\beta]}i^{[\gamma]}}$ are matrices of zeros.
\end{definition}

\begin{figure}
\begin{center}
\centering 
\includegraphics[scale =0.6]{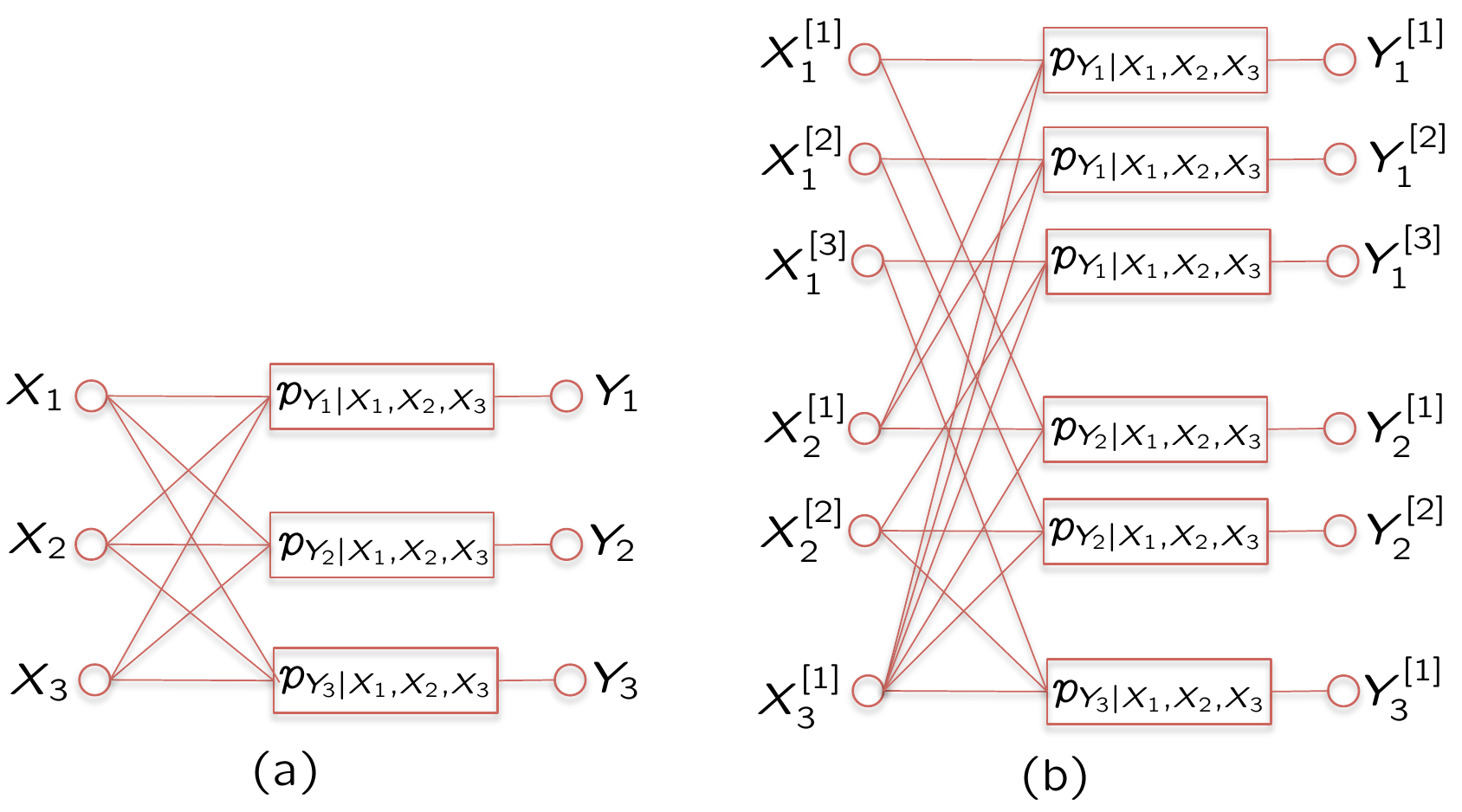}
\caption{(a) A 3-user original interference channel. $p_{Y_i | X_1, X_2, X_3}$ denotes the conditional probability relating the output and input. %transition matrix. 
(b) One possible replicated network when $(\mu_1, \mu_2, \mu_3) = (3, 2, 1)$. The network connectivity is shown in the figure. For example, $Y_2^{[2]}$ is connected to $X_1^{[2]}, X_2^{[2]}$ and $X_3^{[1]}$. The conditional probability relating the output and input is also shown, e.g., $p_{Y_2^{[2]} | X_1^{[2]}, X_2^{[2]}, X_3^{[1]}} = p_{Y_2 | X_1, X_2, X_3}$. \vspace{-0.3cm}}\label{weighted}
\end{center}
\end{figure}

In words, we require that in the replicated network, each desired link is the same as that of the original network, and each replicated receiver sees $K-1$ interferences, one from each interfering replicated transmitter. For a pictorial illustration of one replicated network for the case where $K = 3, (\mu_1, \mu_2, \mu_3) = (3, 2 ,1)$, see Fig. \ref{weighted}. Note that to highlight the generality of the replicated network, we draw the example in the discrete memoryless channel setting.

In the replicated network, each transmitter has the same power constraint as that of the original network and the Gaussian noise at each receiver has the same covariance matrix as that of the original network. Each transmitter has an independent message for its desired receiver. The replicated network is constructed so that its sum capacity is an outer bound to the weighted sum rate of the original network. We state this result in the following theorem.

\begin{theorem}\label{red}
For an integer weight vector $(\mu_1, \cdots, \mu_K)$, the weighted sum rate $\mu_1 R_1 + \cdots \mu_K R_k$ of the original network is bounded by the sum capacity ${\hat R}_\Sigma$ of the replicated network.
\end{theorem}

\proof We show that $\mu_1 R_1 + \cdots \mu_K R_k \leq {\hat R}_\Sigma$. It suffices to prove that if the rate tuple $(R_1, \cdots, R_K)$ is achievable over the original network, then the rate tuple $$(\underbrace{R_1, \cdots, R_1}_{\mu_1 ~\mbox{times}}, \underbrace{R_2, \cdots, R_2}_{\mu_2 ~\mbox{times}}, \cdots, R_K)$$ is achievable over the replicated network.  This is proved by using the encoding/decoding mappings of the original network in the replicated network. Suppose we are given a sequence of encoding and decoding mappings such that  $(R_1, \cdots, R_K)$ is achievable over the original network. Then each Replicated Transmitter $i^{[\alpha]}$ encodes its desired message with the same encoding function as used by Transmitter $i$ in the original network. As a result, from our construction of the replicated network, the received signal at each Replicated Receiver $j^{[\beta]}$ is statistically the same as the received signal at Receiver $j$ in the original network, such that the same decoding mapping can be used to achieve the same rate $R_j$. Therefore the proof is complete. \hfill\QED

{\it Remark:} It is not hard to see that the replicated network argument not only applies to Gaussian channels, but also to discrete memoryless channels. In this work we focus only on Gaussian channels and leave the extension to discrete memoryless channels as future work.

The above theorem is proved in terms of capacity, such that corresponding result on DoF is directly implied. Next we focus on sum-DoF (i.e., choose $\mu_k$ to be the same for all $k$) of the original  $(M_k\times N_k)$ network, which can be bounded in terms of the sum-DoF of the replicated network. 

Although the sum-DoF outer bound problem has been reduced to the sum-DoF outer bound problem of the replicated network, the latter is not available immediately. To obtain an explicit thus easily applicable bound on the sum-DoF of the replicated network, we turn to a simple cooperation based argument. Somewhat surprisingly, a simple cooperation argument for the replicated network can provide tighter bound than possible through the same simple cooperation argument for the original network. In this work, we only apply the simple cooperation argument to bound the replicated network and leave more sophisticated methods and full potential of using the replicated network as future work. 

We use the cooperation argument in the following way. Assume that we replicate each user $\mu$ times. For the resulting $K\mu$-user interference channel, we divide the users into two groups and allow full cooperation between the transmitters/receivers in each group. Thus, we have a 2-user interference channel. For such a 2-user channel, we denote the number of antennas at Transmitter 1 and Receiver 1 as ${\bar M}_1$ and ${\bar N}_1$, respectively. Similarly, we denote the number of antennas at Transmitter 2 and Receiver 2 as ${\bar M}_2$ and ${\bar N}_2$, respectively. The ${\bar N}_2 \times {\bar M}_1$ channel matrix from Transmitter 1 to Receiver 2 is represented as $\bar{\bf H}^{\mbox{\tiny coop}}$. We are now ready to state the outer bound for the  $(M_k\times N_k)$ channel, in the following theorem.

\begin{theorem}\label{theorem:general_k}
{\it For arbitrary realizations of the rank-constrained $K$-user ($M_k\times N_k$) MIMO interference channel, the sum-DoF value is outer bounded as follows.
\begin{equation}
%\mbox{DoF}
d_\Sigma
\le\frac{1}{\mu}\left[ \bar{M}_1 + \bar{N}_2 - \mathrm{rank}\,(\bar{\bf H}^{\mbox{\tiny coop}})\right], \ \ \forall \mu \in\mathbb{Z}^+.
%d_\Sigma
%\le\frac{1}{m+n}\left[m M_\Sigma+n N_\Sigma-\mathrm{rank}\,({\bf \bar{H}}_{n N_\Sigma\times m M_\Sigma})\right], \ \ \forall m,\ n\in\mathbb{Z}^+.
\end{equation}}
where $\bar{\bf H}^{\mbox{\tiny coop}}$ is the interference channel in the replicated network after cooperation, as defined above.
%the cooperation interference matrix in Definition \ref{def::Interference_Matrix}. 
\end{theorem}

{\it Remark: } For the same $\mu$, there may be multiple possible replicated networks. For each possible replicated network, we also have multiple choices of forming groups (cooperation). $\bar{\bf H}^{\mbox{\tiny coop}}$ is defined according to one specific grouping of one specific replicated network. In this regard, Theorem \ref{theorem:LDoF} is a special case of Theorem \ref{theorem:general_k} and it corresponds to the case where $\mu = 2$, the interference links in the replicated network are all between the two replicas of the original network, and cooperation is allowed within each replica of the original network.

\proof 
By Theorem \ref{red}, the sum-DoF value $d_\Sigma = \frac{1}{\mu} (\mu d_1 + \cdots + \mu d_K)$ of the original network is bounded by $\frac{1}{\mu}$ of the sum-DoF of the replicated network, which is in turn bounded by $\frac{1}{\mu}$ of the sum-DoF of the 2-user interference channel after cooperation. Then by Theorem 2 in \cite{krishnamurthy2015degrees}, the sum-DoF of the 2-user channel is bounded by $\bar{M}_1 + \bar{N}_2 - \mathrm{rank}\,(\bar{\bf H}^{\mbox{\tiny coop}})$, and the proof follows. \hfill\QED

As there are multiple choices of replicated networks, it could be computationally cumbersome to find the one that would produce the tightest outer bound. Remarkably, if we relax our target from information theoretic DoF outer bounds to linear DoF outer bounds (i.e., the highest DoF achievable through linear precoding schemes), then a simpler alternative presents itself. 

To present the result, we will need the following definition.
\begin{definition} \label{created}
Suppose we have an $(M_k \times N_k)$ channel with channel matrices ${\bf H}_{ji}$. Similar to Definition \ref{def::new_network}, we replicate  User $i$ $\mu_i$ times. We use notations with tilde symbol in this created network. The channels in the new network are designed as follows.

1) $\forall i, \alpha, \tilde{\bf H}_{i^{[\alpha]}i^{[\alpha]}}= {\bf H}_{ii}$, and $\forall \gamma \in \mathcal{I}_{\mu_i}, \gamma \neq \alpha, \tilde{\bf H}_{i^{[\gamma]}i^{[\alpha]}}$ are matrices of zeros,
%, and $\forall \gamma \in \mathcal{I}_{\mu_i} \neq \alpha$, $\hat{\bf H}_{i^{[\alpha]}i^{[\beta]}}$ is the matrix of zeros, 

2) $\forall i \neq j$, $\forall \beta, \alpha$, $\tilde{\bf H}_{j^{[\beta]}i^{[\alpha]}}=a_{j^{[\beta]}i^{[\alpha]}} {\bf H}_{ji}$, where $a_{j^{[\beta]}i^{[\alpha]}}$ is independently and uniformly drawn from the interval $[0,1]$.% and $\forall \gamma \in \mathcal{I}_{\mu_i}$, $\gamma \neq \alpha$, $\hat{\bf H}_{j^{[\beta]}i^{[\gamma]}}$ are matrices of zeros.
\end{definition}

In words, each replicated receiver here is connected to all interfering replicated transmitters, instead of seeing only one interference from each replicated transmitter, as in the replicated network. As such, in this new network, each receiver is connected to more transmitters than that of the original network, such that the decoding mapping used by the original network does not apply to the new network. Therefore, the new network does not serve as outer bound to the original network information theoretically, but we show that the outer bounding argument still holds in linear sense. We state the result in the following theorem.

\begin{theorem}
For an integer weight vector $(\mu_1, \cdots, \mu_K)$, if the DoF tuple $(d_1, \cdots, d_K)$ is linearly achievable over the original network, then the DoF tuple $(\underbrace{d_1, \cdots, d_1}_{\mu_1 ~\mbox{times}}, \underbrace{d_2, \cdots, d_2}_{\mu_2 ~\mbox{times}}, \cdots, d_K)$ is also achievable linearly over the created network defined in Definition \ref{created}.
\end{theorem}

\proof
As the DoF tuple $(d_1, \cdots, d_K)$ is linearly achievable over the original network, we have integers $m_k, n_k$ such that $d_k = \frac{m_k}{n}$ and User $k$ is able to send $m_k$ symbols with $n$ channel users through linear beamforming schemes. This means that there exist $K$ beamforming matrices ${\bf V}_k \in \mathbb{C}^{M_k n \times m_k}$ used by each transmitter, respectively, and $K$ filtering matrices ${\bf U}_k \in \mathbb{C}^{m_k \times N_k n}$ used by each receiver, respectively, such that
\begin{eqnarray} \label{linearworks}
 \mathrm{rank} ( {\bf U}_k {\bf H}_{kk}^{\mbox{\tiny ex}} {\bf V}_k) = m_k \label{linear1} \\
 {\bf U}_j {\bf H}_{ji}^{\mbox{\tiny ex}}  {\bf V}_i = {\bf 0}, \forall j \neq i \label{linear2}
\end{eqnarray}
where ${\bf H}_{ji}^{\mbox{\tiny ex}}$ denotes the block diagonal channel matrix with $n$ blocks and each block is ${\bf H}_{ji}$. We now proceed to show that User $k^{[\gamma]}, \gamma \in \mathcal{I}_{\mu_k}$ in the created network can also send $m_k$ symbols over $n_k$ channel uses, such that $d_k$ DoF are achievable. For such a purpose, Transmitter $k^{[\gamma]}$ precodes its desired symbols through the beamforming matrix ${\bf V}_k$ and Receiver $k^{[\gamma]}$ decodes its desired symbols with the filtering matrix ${\bf U}_k$. As the desired channel matrices and interference channel matrices (although the number has increased) are the same as that of the original network, from (\ref{linear1}) (\ref{linear2}), all desired symbols can be decoded successfully. This completes the proof. \hfill\QED

{\it Remark:} For a given weight vector, while the replicated network for information theoretic DoF bounds is not unique, the created network for linear DoF bounds is unique. This makes it much easier to explore
the linear DoF outer bound. To find explicit DoF bound on the created network, which serves as outer bound to the linear DoF of the original network, we may also resort to simple cooperation arguments.

\section{Recovering Prior Results as Special Cases}\label{recover}
In this section, we will show that, the prior results in \cite{krishnamurthy2015degrees,liu2014dof} on the optimality of half-the-cake can be recovered as special cases of Theorem \ref{theorem:main}.
\subsection{Full rank case}
In \cite{liu2014dof}, half-the-cake DoF is shown to be optimal in a $K$-user $M_k\times M_k$ MIMO interference channel where there is no dominant user and all channels have full rank. To prove that  full rank $K$-user $M_k\times M_k$ MIMO interference channels  satisfy the condition in Theorem \ref{theorem:main}, it is sufficient to show that for any $M_1\le M_2+\cdots+M_K$, we can always find  a set of values for $\bar{D}_{ij}\leq \min(M_i,M_j)$ that satisfy the condition in Lemma 1. 

To start, suppose $\forall k\in\mathcal{I}_K$, each Transmitter $k$  has $M_k$ chips and each Receiver $k$ has an empty bin that can hold $M_k$ chips. Transmitter 1 starts by dropping as many chips as possible into Receiver 2's bin, and then if the bin is full and he still has chips left over, he continues with Receiver 3's bin, and so on. After Transmitter 1 is done, Transmitter 2 does the same, starting with Receiver 3's bin. Transmitter 2 is followed by Transmitters $3, 4, \cdots, K$, in that order. At the end, the number of chips in receiver bin $i$ from Transmitter $j$ is chosen to be the rank $\bar{D}_{ij}$. Since there is no dominant user, the total capacity of all bins is the same as the total number of chips, and users are arranged as $M_1\geq M_2\geq \cdots\geq M_K$, it is easy to see that this allocation works.

\subsection{Symmetric case}
In \cite{krishnamurthy2015degrees}, it is shown that for a $K$-user rank deficient MIMO interference channel with $M$ antennas at each node, if all the direct channels have full rank, and all cross channels have rank $D$, then half-the-cake DoF is optimal when $(K-1)D\ge M$. We now show that this result is also a special case of Theorem \ref{theorem:main}.

Note that if $\tfrac{M}{K-1}$ is an integer, then we just need to reduce $D$ to the value $\tfrac{M}{K-1}$. When $\tfrac{M}{K-1}$ is not an integer, we can write $M=\left \lfloor \tfrac{M}{K-1} \right \rfloor (K-1)+\Delta$ for some positive integer $\Delta<K-1$. Now, assign reduced interference ranks as follows.
\begin{eqnarray*}
\bar{D}_{ji}&=&\left \lfloor \tfrac{M}{K-1} \right \rfloor + 1  \leq D, \mbox{ if } j\in\{i+1, i+2, \cdots, i+\Delta\},\\
\bar{D}_{ji}&=&\left \lfloor \tfrac{M}{K-1} \right \rfloor \leq D, \mbox{ otherwise}.
\end{eqnarray*}
With these reduced ranks, the condition in Lemma 1 is always satisfied. Thus, Theorem \ref{theorem:main} applies and half-the-cake DoF is optimal.

\section{Examples of Applications of New Outer Bounds}\label{sec:two_example}

As an example of the broader applicability of the new DoF outer bounds, we next recover a known DoF result in $(M\times N)$ setting with our new bound. After that, we will apply the new bound to the generalized $(M_k\times N_k)$ channel.

\subsection{Example 1: $(M\times N)$ Channel}\label{sec:mnexample}
We consider a 3-user $(2\times3)$ generic full rank MIMO interference channel. It is shown in \cite{wang2014subspace} that the sum-DoF value of this channel is $\tfrac{18}{5}$. We will show that the sum-DoF outer bound can be obtained in a simple manner by using a replicated network and cooperation based bound. 

The replicated network is described as follows.
We set $\mu_1 = \mu_2 = \mu_3 = \mu = 5$, i.e., we replicate each user $i \in \mathcal{I}_3$ 5 times. The desired channels in the replicated network are the same as that in the original network.
%, i.e., $\alpha \in \mathcal{I}_5, \hat{\bf H}_{i^{[\alpha]}i^{[\alpha]}}= {\bf H}_{ii}$. 
The interference channels are chosen as, $\forall \alpha, \beta \in \mathcal{I}_5, \hat{\bf H}_{i^{[\beta]} {(i+1)}^{[\alpha]}}$ equals ${\bf H}_{i (i+1)}$ if $\alpha = \beta - 3$, and ${\bf 0}_{3 \times 2}$ otherwise, %if $\alpha \neq \beta - 3$, 
$\hat{\bf H}_{i^{[\beta]} {(i+2)}^{[\alpha]}}$ equals ${\bf H}_{i (i+2)}$ if $\alpha = \beta - 2$, and ${\bf 0}_{3 \times 2}$ otherwise. %if $\alpha \neq \beta - 2$.
It can be verified that the replicated network satisfies Definition \ref{def::new_network}.

Next we allow the first three replicas of the original network (i.e., Users $1^{[l]}, 2^{[l]}, 3^{[l]}, l = 1,2, 3$) to cooperate, %and form group 2, 
and the remaining users to cooperate. %and form group 1. 
This creates a 2-user interference channel where Transmitter 1 has ${\bar M}_1 = 18$ antennas, Receiver 1 has ${\bar N}_1 = 27$ antennas, Transmitter 2 has $\bar{M}_2 = 12$ antennas and Receiver 2 has $\bar{N}_2 = 18$ antennas. $\bar{\bf H}^{\mbox{\tiny coop}}$ is the $18\times18$ interference matrix  from Transmitter 1 to Receiver 2.
By Theorem \ref{theorem:general_k}, we have $d_\Sigma
\le\frac{1}{\mu}\left[ \bar{M}_1 + \bar{N}_2 - \mathrm{rank}\,(\bar{\bf H}^{\mbox{\tiny coop}})\right] = \frac{1}{5} [18 + 18 - \mathrm{rank}\,(\bar{\bf H}^{\mbox{\tiny coop}})]$. In order to prove $\frac{18}{5}$ is a valid outer bound, we are left to prove that $\mathrm{rank}\,(\bar{\bf H}^{\mbox{\tiny coop}}) = 18$, i.e., $\bar{\bf H}^{\mbox{\tiny coop}}$ has full rank almost surely.

To show this, it suffices to prove that the determinant polynomial of $\bar{\bf H}^{\mbox{\tiny coop}}$ is not identically zero, which can be proved by constructing a specific channel such that this is true. One such channel may be
\begin{equation}
{\bf H}_{i(i+1)}%={\bf H}_{23}={\bf H}_{31}
=\begin{bmatrix}
1& 0    \\
0& 1\\
0&0
\end{bmatrix},
%\end{equation}
%\begin{equation}
{\bf H}_{i(i+2)}%={\bf H}_{21}={\bf H}_{32}
=\begin{bmatrix}
0& 0    \\
1& 0\\
0&1
\end{bmatrix}.
\end{equation}
It is readily verifiable that for such a channel, the determinant of $\bar{\bf H}^{\mbox{\tiny coop}}$ is non-zero. Therefore, $\bar{\bf H}^{\mbox{\tiny coop}}$ has full rank and the proof is complete.

\subsection{Example 2: $(M_k\times N_k)$ channel}
We now consider a 3-user $(10 \times 10)(8 \times 10)(6 \times 3)$  MIMO interference channel. %with $(M_1, M_2, M_3; N_1, N_2, N_3)=(10, 8, 6; 10, $ $10, 3)$. 
It is assumed that ${\bf H}_{31}$ is the matrix of zeros, i.e., $D_{31}=0$, and all other interference matrices are generic full rank. This channel setting has not been considered in the literature and its sum-DoF value is not known. We show that the sum-DoF value is 12, with the help of the insights from our general outer bound (Theorem \ref{theorem:general_k}).

We start with the outer bound. 
%Choose The sum-DoF outer bound is 12 according to Theorem \ref{theorem:general_k} by letting $\mu=2$. 
The replicated network is described as follows. We set $\mu_1=\mu_2=\mu=2$, i.e., we replicate each user 2 times. The channels in the replicated network are chosen so that $ \forall i,j\in\mathcal{I}_3$,  1) $\hat{\bf H}_{j^{[1]}i^{[2]}}=\hat{\bf H}_{j^{[2]}i^{[1]}}= {\bf H}_{ji}$ whenever $i\neq j$, 2) $ \hat{\bf H}_{i^{[1]}i^{[1]}}=\hat {\bf H}_{i^{[2]}i^{[2]}}= {\bf H}_{ii}$, 3) $ \hat{\bf H}_{j^{[1]}i^{[1]}}=\hat{\bf H}_{j^{[2]}i^{[2]}}$ is the matrix of zeros whenever $i\neq j$, and 4) $\hat{\bf H}_{i^{[1]}i^{[2]}}=\hat{\bf H}_{i^{[2]}i^{[1]}}$ is the matrix of zeros. It can be verified that this replicated network satisfies Definition \ref{def::new_network}.

Next we allow users $1^{[1]}$, $2^{[1]}$ and $3^{[1]}$ to cooperate, and users $1^{[2]}$, $2^{[2]}$ and $3^{[2]}$ to cooperate. This creates a 2-user interference channel where Transmitter 1 has ${\bar M}_1 = 24$ antennas, Receiver 1 has ${\bar N}_1 = 23$ antennas, Transmitter 2 has $\bar{M}_2 = 24$ antennas and Receiver 2 has $\bar{N}_2 = 23$ antennas. $\bar{\bf H}^{\mbox{\tiny coop}}$ is the $23\times24$ interference matrix  from Transmitter 1 to Receiver 2. If $\bar{\bf H}^{\mbox{\tiny coop}}$ has full rank, then by Theorem \ref{theorem:general_k}, we have the desired result, $d_\Sigma\le\frac{1}{\mu}\left[{\bar M}_1+\bar{N}_2-\mathrm{rank}\,(\bar{\bf H}^{\mbox{\tiny coop}})\right]=\frac{1}{2}\left[24+23-23\right]=12$.

To show that $\bar{\bf H}^{\mbox{\tiny coop}}$ has full rank almost surely, it suffices to prove that the determinant polynomial of $\bar{\bf H}^{\mbox{\tiny coop}}$ is not identically zero, which can be proved by constructing a specific channel such that this is true. One such channel may be
\begin{align*}
{\bf H}_{21}=&{\bf I}_{N_2},\ \ {\bf H}_{32}=\begin{bmatrix}{\bf I}_{N_3}&{\bf 0}_{N_3\times(M_2-N_3)}\end{bmatrix},\\
{\bf H}_{13}=&{\bf H}_{23}=\begin{bmatrix}{\bf I}_{M_3}\\{\bf 0}_{(N_1-M_3)\times{M_3}}\end{bmatrix},\\
{\bf H}_{12}=&\begin{bmatrix}{\bf 0}_{(N_1-M_2)\times{M_2}}\\{\bf I}_{M_2}\end{bmatrix}.
\end{align*}
It is readily verifiable that for such a channel, the determinant of $\bar{\bf H}^{\mbox{\tiny coop}}$ is non-zero. Therefore, $\bar{\bf H}^{\mbox{\tiny coop}}$ has full rank and the outer bound proof is complete.

We next proceed to the achievability. We show that %this outer bound is tight by verifying that 
the DoF tuple $(d_1, d_2, d_3)=(7, 3, 2)$ can be achieved, such that the sum-DoF bound of $12$ is tight. We use ${\bf v}_{i1}$, ${\bf v}_{i2}$, ..., ${\bf v}_{id_i}$ to denote the beamforming vectors at Transmitter $i$. 
We first choose ${\bf v}_{21}$ and ${\bf v}_{31}$ so that 
\begin{align} \label{cc1}
{\bf H}_{32} {\bf v}_{21} ={\bf 0}, {\bf H}_{12} {\bf v}_{21} ={\bf H}_{13} {\bf v}_{31}  ~~\Leftrightarrow~~~
%\end{align}
%\begin{align}
\begin{matrix}\underbrace{\begin{bmatrix}
\mathbf{H}_{32} & \mathbf{0}      \\
\mathbf{H}_{12} & -\mathbf{H}_{13}
\end{bmatrix}_{13\times 14}}\\ \mathbf{A}\end{matrix}
\begin{matrix}\underbrace{\begin{bmatrix}
{\bf v}_{21}   \\
{\bf v}_{31}
\end{bmatrix}}\\ \mathbf{v}\end{matrix}= \mathbf{0}. 
\end{align}
Note that $\mathbf{v}$ can be chosen from the right null space of $\mathbf{A}$. 

Next we choose ${\bf v}_{22}$ so that
\begin{align}
{\bf H}_{32}  {\bf v}_{22} = 0, ~\mbox{and}~ {\bf v}_{22} ~\mbox{is linearly independent of}~ {\bf v}_{21}, \label{cc2}
\end{align}

and ${\bf v}_{23}, {\bf v}_{32}$ so that 
\begin{align}
{\bf H}_{12} {\bf v}_{23} ={\bf H}_{13} {\bf v}_{32}, ~\mbox{where}~ {\bf v}_{23} ~\mbox{is linearly independent of}~ {\bf v}_{21}, {\bf v}_{22}, ~\mbox{and}~ {\bf v}_{32} ~\mbox{is independent of}~ {\bf v}_{31}. \label{cc3}
\end{align}

The existence of ${\bf v}_{22}$ is guaranteed as ${\bf H}_{32}$ is a $3 \times 8$ generic matrix, whose right null space has 5 dimensions. The existence of ${\bf v}_{23}, {\bf v}_{32}$ is guaranteed as ${\bf H}_{12}$ has dimension $10 \times 8$ and ${\bf H}_{13}$ has dimension $10 \times 6$, such that the two overlap in a 4 dimensional subspace.

Then we choose ${\bf v}_{11}, {\bf v}_{12}$ so that 
\begin{align}
&{\bf H}_{21}\begin{bmatrix}{\bf v}_{11}&{\bf v}_{12}\end{bmatrix}={\bf H}_{23}\begin{bmatrix}{\bf v}_{31}&{\bf v}_{32}\end{bmatrix} \Rightarrow \begin{bmatrix}{\bf v}_{11}&{\bf v}_{12}\end{bmatrix} = {\bf H}_{21}^{-1} {\bf H}_{23}\begin{bmatrix}{\bf v}_{31}&{\bf v}_{32}\end{bmatrix} \label{cc4}
\end{align}

At the last step, ${\bf v}_{13}$, ..., ${\bf v}_{17}$ are chosen as generic vectors. Thus, we have allocated all the beamforming vectors.

We are left to show that at each receiver, the interferences are aligned to a subspace that is independent of the desired signal space. First, we consider Receiver 3. Note that ${\bf H}_{31} = 0$. From (\ref{cc1}) (\ref{cc2}), the interference space is ${\bf H}_{32} [{\bf v}_{21}, {\bf v}_{22}, {\bf v}_{23} ] = {\bf H}_{32} {\bf v}_{23}$, which has dimension $1 = N_3 - d_3$. Next, we consider Receiver 2. From (\ref{cc4}), the interference from Transmitter 3 lies in the span of the interference from Transmitter 1, so that the total interference occupies $d_1 = 7$ dimensions, leaving $10 -7 = 3 = d_2$ dimensions for the desired signal, as desired. 
We now consider Receiver 1. From (\ref{cc2}) (\ref{cc3}), the interference from Transmitter 3 lies in the span of the interference from Transmitter 2, so that the total interference has $d_3 = 3$ dimensions. The desired signal is left with $10 - 3 = 7 = d_1$ dimensions. 
Finally, as desired channels do not appear when we design the beamforming vectors, the independence of the aligned interference and desired signal is guaranteed. This completes the proof.

\section{Conclusion}
The motivation for this work was to explore the sharper insights, especially into information theoretic DoF outer bounds, that might emerge from the study of rank-deficient MIMO interference channels under a model that unifies and generalizes prior works. For a $K$-user MIMO interference channel with arbitrarily rank-deficient cross-channels, where there are $M_k$ antennas at the $k^{th}$ user pair, it was shown that the sum-DoF cannot exceed half-the-cake if the overall $M_\Sigma\times M_\Sigma$ channel matrix $\bar{\bf H}$ where all desired channels have been set to zero, has full rank. This was accomplished through a new outer bound based on the idea of creating a replicated-network, i.e., creating copies (replicas) of certain users and choosing the connectivity of the replicated network in such a way that any achievable scheme in the original network translates into an achievable scheme for the replicated network. Depending on the number of replicas created for each user, the sum rate of the replicated network bounds the corresponding weighted sum of rates from the original network. What is remarkable about the replicated network is that it creates a new perspective of the problem, so that even simple arguments such as user cooperation become quite powerful when applied in the replicated network, giving rise to stronger outer bounds, than when applied directly in the original network. The replication argument is applicable not only to arbitrary MIMO interference channels with arbitrary rank-constraints, but much more broadly, even beyond Gaussian interference channels. The conceptual simplicity and apparent breadth of replication based bounds calls for future work into understanding their full potential, especially  for MIMO interference channels where the DoF remain open in general.

\appendix
\section*{Appendix}

\section{Counterexample to Original Insight}\label{sec:example}
\begin{figure}
%\begin{center}
\centering 
\includegraphics[scale =0.4]{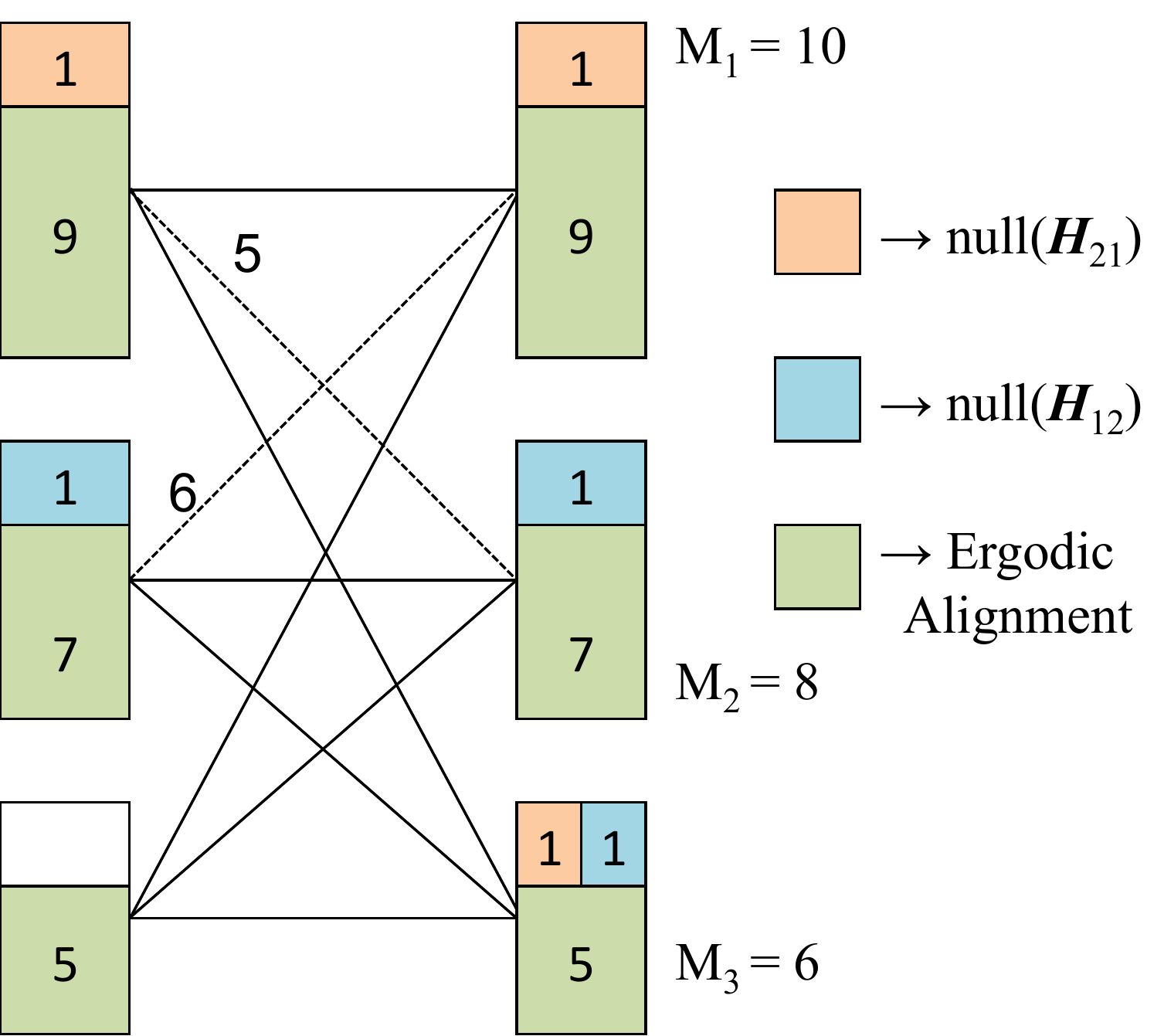}
\caption{Example for achieving more than half-the-cake DoF.\vspace{-0.3cm}}\label{example}
%\end{center}
\end{figure}

Here we briefly summarize how more than half-the-cake DoF can be achieved in  the 3-user setting shown in Fig. \ref{example}, where $D_{12}=6$, $D_{21}=5$ and all other links have full rank. 

The transmission takes place over 2 channel uses, where all cross channels remain the same, and all direct channels change to different generic values \cite{Jafar_BlindIA}. We use $\mathbf{v}_{1}^{z}$ and $\mathbf{v}_{2}^{z}$ to denote the beamforming vectors at Transmitters 1 and 2 that need to be aligned at Receiver 3 after being chosen from the null space they see at each other. The symbols carried by $\mathbf{v}_{1}^{z}$ and $\mathbf{v}_{2}^{z}$ are different over two channel uses. Mathematically, we have
%\begin{align*}
%\mathbf{H}_{21}\mathbf{V}_{1}^{z}&=\mathbf{0}\\
%\mathbf{H}_{12}\mathbf{V}_{2}^{z}&=\mathbf{0}\\
%\mathbf{H}_{31}\mathbf{V}_{1}^{z}&=\mathbf{H}_{32}\mathbf{V}_{2}^{z}
%\end{align*}
%$$
%\Rightarrow
%\begin{matrix}\underbrace{\begin{bmatrix}
%\mathbf{H}_{21} & \mathbf{0}      \\
%\mathbf{0} & \mathbf{H}_{12} \\
%\mathbf{H}_{31} & -\mathbf{H}_{32}
%\end{bmatrix}_{24\times 18}}\\ \mathbf{A}\end{matrix}
%\begin{matrix}\underbrace{\begin{bmatrix}
%\mathbf{V}_{1}^{z}   \\
%\mathbf{V}_{2}^{z}
%\end{bmatrix}}\\ \mathbf{v}\end{matrix}= \mathbf{0}. 
%$$
$$
\begin{matrix}
\mathbf{H}_{21}\mathbf{v}_{1}^{z}=\mathbf{0},\\
\mathbf{H}_{12}\mathbf{v}_{2}^{z}=\mathbf{0},\\
\mathbf{H}_{31}\mathbf{v}_{1}^{z}=\mathbf{H}_{32}\mathbf{v}_{2}^{z}.
\end{matrix}
\Rightarrow
\begin{matrix}\underbrace{\begin{bmatrix}
\mathbf{H}_{21} & \mathbf{0}      \\
\mathbf{0} & \mathbf{H}_{12} \\
\mathbf{H}_{31} & -\mathbf{H}_{32}
\end{bmatrix}_{24\times 18}}\\ \mathbf{A}\end{matrix}
\begin{matrix}\underbrace{\begin{bmatrix}
\mathbf{v}_{1}^{z}   \\
\mathbf{v}_{2}^{z}
\end{bmatrix}}\\ \mathbf{v}\end{matrix}= \mathbf{0}. 
$$
Note that matrix $\mathbf{A}$ has rank 17, thus $\mathbf{v}$ can be chosen from the right null space of $\mathbf{A}$. In the same manner, we choose the receive combining vectors $\mathbf{u}_{1}^{z}$ and $\mathbf{u}_{2}^{z}$ at Receivers 1 and 2 satisfying the following equations
$$
\mathbf{u}_{1}^{z}\mathbf{H}_{12}=\mathbf{0},\quad\mathbf{u}_{2}^{z}\mathbf{H}_{21}=\mathbf{0},\quad\mathbf{u}_{1}^{z}\mathbf{H}_{13}=\mathbf{u}_{2}^{z}\mathbf{H}_{23}.
$$

Next, we use $\mathbf{V}_{k}^{e}$ and $\mathbf{U}_{k}^{e}$ to denote the $M_k\times (M_k-1)$ and $(M_k-1)\times M_k$  matrices at each transmitter and receiver, respectively. These matrices carry the signals for ergodic alignment (green area in Fig. \ref{example}), i.e., signals repeated over the two channel uses. User 3 needs to choose its beamforming/combining matrices to satisfy  $\mathbf{V}_{3}^{e}=\mathrm{span}(\mathrm{null}(\mathbf{u}_{2}^{z}\mathbf{H}_{23}))$ and $\mathbf{U}_{3}^{e}=\mathrm{span}(\mathrm{null}(\mathbf{H}_{32}\mathbf{v}_{2}^{z}))$.
As a result, each receiver can eliminate interference by only subtracting the part of received signals corresponding to $\mathbf{U}_{k}^{e}$ of two time slots. Thus, a total of 25 DoF are achieved over the two channel uses, or equivalently, 12.5 DoF per channel use (half-the-cake is 12 DoF per channel use).

\section{Proof of Lemma \ref{lemma:connect}}\label{proof:connect}
We prove Lemma \ref{lemma:connect} by first showing that  Condition \eqref{lemma_eq} is sufficient for $\bar{\mathbf{H}}$ to have full rank, and then showing that this condition is also necessary.
\subsection{Sufficiency}
\begin{figure}
%\begin{center}
\centering 
\includegraphics[scale =0.7]{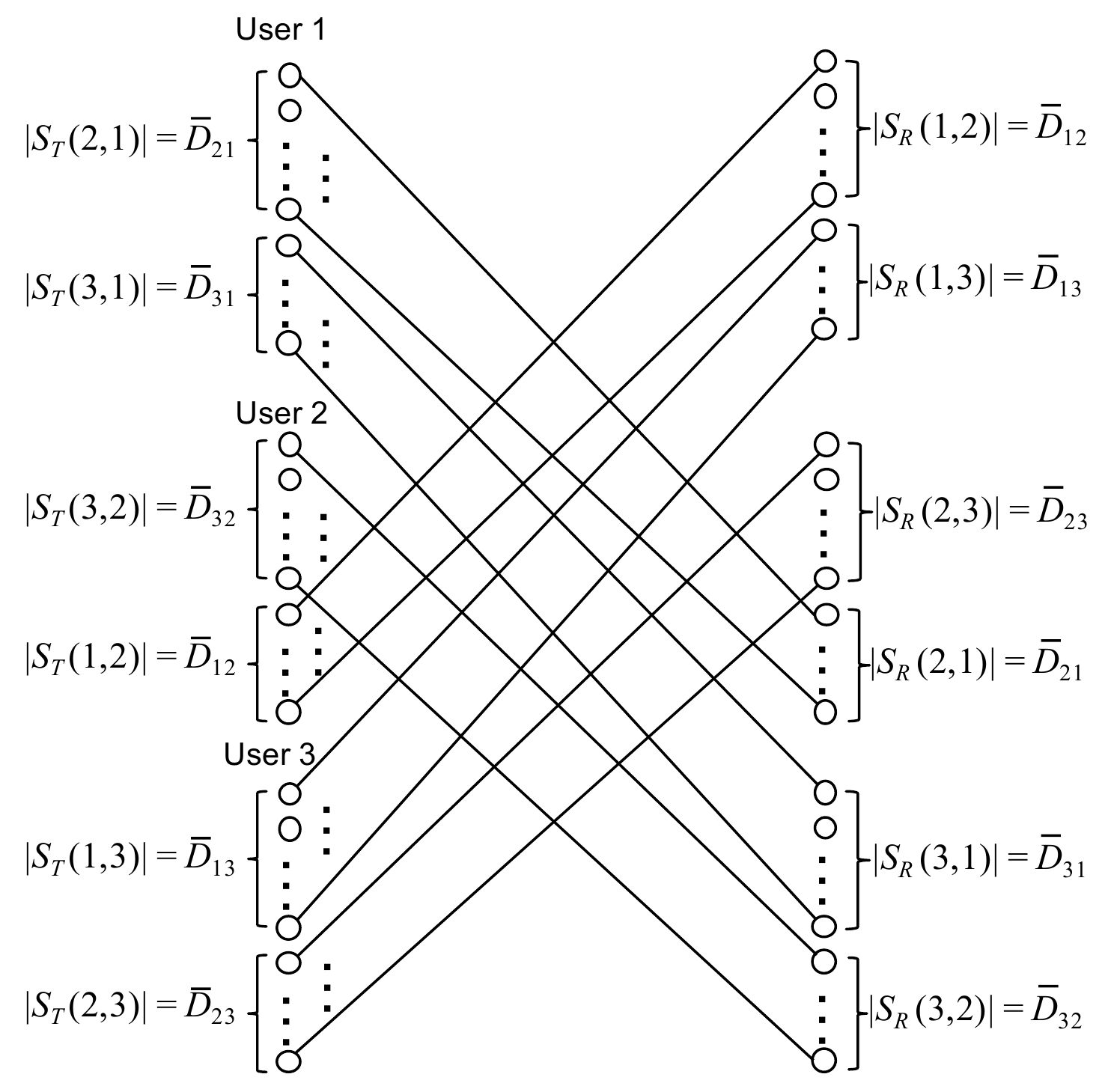}
\caption{Illustration of the sufficiency proof for Lemma \ref{lemma:connect} when $K=3$.\vspace{-0.3cm}}\label{fig_b2}
%\end{center}
\end{figure}

To prove that $\bar{\bf H}$ is full-rank almost surely for generic rank-deficient channels with given ranks, it suffices to show that its determinant polynomial is not  identically zero. To show this, it suffices to find one realization of ${\bar{\bf H}}$ for which the determinant is not zero. Such a realization is  constructed as follows. At Receiver $i$, starting from the first antenna, label the first set of $\bar{D}_{i,i+1}$ antennas as $S_R(i,i+1)$, the next  $\bar{D}_{i,i+2}$ as $S_R(i,i+2)$, and so on, until the final set of $\bar{D}_{i,i+K-1}$ antennas is labeled as $S_R(i,i+K-1)$. Similarly, at Transmitter $j$, starting from the first antenna, label the first set of $\bar{D}_{j+1,j}$ antennas as $S_T(j+1,j)$, the next set of $\bar{D}_{j+2,j}$ antennas as the set $S_T(j+2,j)$, and so on until the last set of $\bar{D}_{j+K-1,j}$ antennas is labeled as $S_T(j+K-1,j)$. Now connect transmit antennas in $S_T(i,j)$ with the receive antennas in $S_R(i,j)$ through identity matrices. For a pictorial illustration of such channel realization for the case where $K = 3$, see Fig. \ref{fig_b2}. With this channel realization, each transmit antenna is connected to exactly one undesired receive antenna, so that $\bar{\bf H}$ has exactly one $1$ in each row and each column, and is therefore full rank. Increasing any of the ranks only introduces additional variables into the polynomial which can be set to zero to return to the same realization described above, thus proving that the polynomial is not identically zero.

\subsection{Necessity}
If the following partitioned matrix $\bar{\bf H}$ has full rank,

\begin{equation}\label{hbar}
{\bf \bar{H}}=\bordermatrix{&\mbox{\tiny $M_1$}&\mbox{\tiny $M_2$}&\mbox{\tiny$\cdots$}&\mbox{\tiny $M_{K-1}$}&\mbox{\tiny $M_{K}$}\cr
\mbox{\tiny $M_1$}&{\bf 0}&{\bf H}_{12}&\cdots&{\bf H}_{1(K-1)}&{\bf H}_{1K}\cr
 \mbox{\tiny $M_2$}&{\bf H}_{21}&{\bf 0}&\cdots&{\bf H}_{2(K-1)}&{\bf H}_{2K}\cr
  \mbox{\tiny\quad$\vdots$}&\vdots&{\vdots}&\ddots&{\vdots}&{\vdots}\cr
 \mbox{\tiny $M_{K}$}&{\bf H}_{K1}&{\bf H}_{K2}&\cdots&{\bf H}_{K(K-1)}&{\bf 0}\cr
 }
\end{equation}
then the first observation is that each column of sub-matrix and each row of sub-matrix must have full rank, i.e., the rank of each sub-matrix must satisfy the following conditions.
\begin{equation}\label{fullrank}
\sum_{j\in\mathcal{I}_{K}\backslash i}D_{ji}\ge{M_i},\ \sum_{j\in\mathcal{I}_{K}\backslash i}D_{ij}\ge{M_i},\ \forall i \in\mathcal{I}_{K}.
\end{equation}

With the help of this observation, the necessity of Condition \eqref{lemma_eq} can be proved as follows. Any sub-matrix ${\bf H}_{ji}$ of rank $D_{ji}$ can be represented as a sum of $D_{ji}$ matrices, each of which has rank 1, i.e.,
\begin{equation}\label{summatrix}
\begin{split}
{\bf H}_{ji}&=a_{ji}^{[1]}{\bf v}_{ji}^{[1]}{\bf u}_{ji}^{[1]}+a_{ji}^{[2]}{\bf v}_{ji}^{[2]}{\bf u}_{ji}^{[2]}+\cdots+a_{ji}^{[D_{ji}]}{\bf v}_{ji}^{[D_{ji}]}{\bf u}_{ji}^{[D_{ji}]}
\end{split}
\end{equation}
where ${\bf v}_{ji}^{[m]}$ and ${\bf u}_{ji}^{[m]}$ are $M_j\times1$ and $1\times M_i$ unit vectors, respectively. Now let us consider the $a_{ji}^{[m]}$ as variables while ${\bf v}_{ji}^{[m]}$ and ${\bf u}_{ji}^{[m]}$  are treated as constants.   After all the ${\bf H}_{ji}$ are represented in the form as \eqref{summatrix}, we use $\mathcal{A}$ to denote the set of all the $a_{ji}^{[m]}$ in ${\bf \bar{H}}$. Then we go through the following steps.
\paragraph{Step 1}
Choose any one of the variables $a_{ji}^{[m]}$ from $\mathcal{A}$. We set this variable $a_{ji}^{[m]}$ to zero, then we have reduced the rank of corresponding sub-matrix ${\bf H}_{ji}$ by 1.
\paragraph{Step 2}
We check the determinant polynomial of ${\bf \bar{H}}$ with the rank-reduced sub-matrix ${\bf H}_{ji}$. If det(${\bf \bar{H}}$) is not the zero polynomial, then ${\bf \bar{H}}$ is full rank almost surely, fix $a_{ji}^{[m]}=0$. If det(${\bf \bar{H}}$) is the zero polynomial, leave $a_{ji}^{[m]}$ as a generic variable. Remove this $a_{ji}^{[m]}$ from the set $\mathcal{A}$.
\paragraph{Step 3}
If the set $\mathcal{A}$ is not an empty set, go back to step 1. If $\mathcal{A}$ is empty, i.e., all $a_{ji}^{[m]}$ have been tested, we now have a situation that the remaining $a_{ji}^{[m]}$ must all be non-zero for ${\bf \bar{H}}$ to have full rank $M_\Sigma$. At this stage, the number of remaining $a_{ji}^{[m]}$ variables for each sub-matrix define the reduced rank value $\bar{D}_{ji}$ for that matrix.

Now, based on the following two facts, it can be claimed that the number of remaining $a_{ji}^{[m]}$ variables cannot be more than $M_\Sigma$.

{\bf Fact 1:} ``\emph{Since setting any $a_{ji}^{[m]}$ to zero will make $\det({\bf \bar{H}})$ the zero polynomial, then it must be true that every remaining $a_{ji}^{[m]}$ variable appears in every term of the polynomial.}"

{\bf Fact 2:} ``\emph{Since each term in $\bf \bar{H}$ is linear in each $a_{ji}^{[m]}$ variable, each term of the determinant polynomial cannot involve more than $M_\Sigma$ of $a_{ji}^{[m]}$ variables.}"

Fact 1 says that every remaining $a_{ji}^{[m]}$ must be present in every term of $\det({\bf \bar{H}})$. Fact 2 says that there cannot be more than $M_\Sigma$ remaining $a_{ji}^{[m]}$ that are presented in any given term of $\det({\bf \bar{H}})$. Thus the two facts imply that the number of remaining $a_{ji}^{[m]}$ cannot be more than $M_\Sigma$, i.e., $\sum_{j\in\mathcal{I}_{K}\backslash i}{\bar D}_{ji}\le{M_i},\ \sum_{j\in\mathcal{I}_{K}\backslash i}{\bar D}_{ij}\le{M_i}$. Since all the $\bar{D}_{ji}$ must also satisfy Condition \eqref{fullrank} in order for $\bf \bar{H}$ to have full rank, all the inequalities in \eqref{fullrank} must take equality. In other words, for any full rank matrix $\bf \bar{H}$, there always exist reduced ranks $\bar{D}_{ji}\le D_{ji}$ which satisfy the Condition \eqref{lemma_eq}. This completes the proof.

\section{Proof of Theorem \ref{theorem:new}}\label{proof:new}
%\section{Proof of the Remark under }
Before proceeding to the proof of of Theorem \ref{theorem:new}, we first prove the remark under Theorem \ref{theorem:main}. That is, we want to prove that the following two polytopes are equivalent.

%\proof
The polytope (denoted as $\bar{\mathcal{D}}^*$) given by Condition (\ref{1}) (when $K=3$)  is the set of tuples $(D_{12}, D_{21}, D_{23}, D_{32}, D_{31}, D_{13}) \in \mathbb{Z}_+^6$ such that there exist $\bar{D}_{ji}\leq D_{ji}$, which satisfy the following constraints.
\allowdisplaybreaks
\begin{eqnarray} 
\bar{D}_{12} + \bar{D}_{13} &=& M_1 \label{dd1} \\
\bar{D}_{21} + \bar{D}_{23} &=& M_2 \label{dd2} \\
\bar{D}_{31} + \bar{D}_{32} &=& M_3 \label{dd3} \\
\bar{D}_{21} + \bar{D}_{31} &=& M_1 \label{dd4} \\
\bar{D}_{12} + \bar{D}_{32} &=& M_2 \label{dd5} \\
\bar{D}_{13} + \bar{D}_{23} &=& M_3 \label{dd6} 
\end{eqnarray} .
%\begin{eqnarray}
%\bar{\mathcal{D}}^* = \{\}
%\end{eqnarray}

The polytope (denoted as ${\mathcal{D}}^*$) given by Condition (\ref{5}) is the set of tuples $(D_{12}, D_{21}, D_{23}, D_{32}, D_{31}, D_{13}) \in \mathbb{Z}_+^6$ defined by the following constraints.
%\allowdisplaybreaks
\begin{eqnarray}
D_{12} + D_{13} &\geq& M_1 \label{cd1} \\
D_{21} + D_{23} &\geq& M_2 \label{cd2}\\
D_{31} + D_{32} &\geq& M_3 \label{cd3} \\
D_{21} + D_{31} &\geq& M_1 \label{cd4} \\
D_{12} + D_{32} &\geq& M_2 \label{cd5} \\
D_{13} + D_{23} &\geq& M_3 \label{cd6} \\
D_{12} + D_{21} &\geq& M_1 + M_2 - M_3 \label{cd7} \\
D_{23} + D_{32} &\geq& M_2 + M_3 - M_1 \label{cd8} \\
D_{13} + D_{31} &\geq& M_1 + M_3 - M_2  \label{cd9}
\end{eqnarray}
The above 9 linear inequalities are obtained by expanding each term in the min expression of (\ref{5}) and rearranging.

Next we prove $\mathcal{D}^* = \bar{\mathcal{D}}^*$ by proving that $\mathcal{D}^*\subseteq\bar{\mathcal{D}}^*$ and $\bar{\mathcal{D}}^*\subseteq{\mathcal{D}}^*$.

{\color{black} $\mathcal{D}^*\subseteq\bar{\mathcal{D}}^*: $ We need to show that if $D_{ji}$ satisfy (\ref{cd1}) to (\ref{cd9}), then we can find $\bar{D}_{ji}\leq D_{ji}$ that satisfy (\ref{dd1}) to (\ref{dd6}). Without loss of generality, we assume 
\begin{eqnarray}
\min(M_3+D_{12}, M_1+D_{23}, M_2+D_{31}) = M_1 + D_{23} \label{cd10}
\end{eqnarray}

We set $\bar{D}_{ji}$ as follows.
\begin{eqnarray} 
\bar{D}_{12}  &=& D_{12} - (D_{12} + M_{3} - M_{1} - D_{23}) = M_1 + D_{23} - M_3\\
\bar{D}_{13} &=& D_{13} - 
%\underbrace{
(D_{13} + D_{23} - M_3)
%}
%_{\geq 0 ~\mbox{\small from}~ (\ref{cd6})} 
= M_3 -  D_{23}\\
\bar{D}_{21}  &=& D_{21} - (D_{21} + D_{23} - M_2) =  M_2 - D_{23} \\
\bar{D}_{23}  &=& {D}_{23} \\
\bar{D}_{31} &=& D_{31} - (D_{31} + M_2 - M_1 - D_{23}) = M_1 + D_{23} -M_2 \\
\bar{D}_{32} &=&  D_{32} - (D_{32} + D_{23} + M_1 - M_2 - M_3) = M_2 + M_3 - M_1 - D_{23}
\end{eqnarray}
It is easy to verify that (\ref{dd1}) to (\ref{dd6}) are satisfied by above assignment. We are left to prove that each difference term is valid, i.e., {\color{black} $0 \leq D_{ji} - \bar{D}_{ji} \leq D_{ji} %\in [0, D_{ji}]
$}. This proof is a simple manipulation of the inequalities (\ref{cd1}) to (\ref{cd10}) and the rank property $0 \leq D_{ji} \leq \min(M_i, M_j)$, thus we omit it.
Therefore this direction is proved.}

$\bar{\mathcal{D}}^*\subseteq{\mathcal{D}}^*:$ We need to show that if there exist $\bar{D}_{ji}\leq D_{ji}$ that satisfy (\ref{dd1}) to (\ref{dd6}), then $D_{ji}$ must satisfy (\ref{cd1}) to (\ref{cd9}). To see this, note that we have
\begin{eqnarray}
(\ref{dd1}) + (\ref{dd2}) - (\ref{dd6}) ~\Rightarrow \bar{D}_{12} + \bar{D}_{21} &=& M_1 + M_2 - M_3 \label{dd7} \\
(\ref{dd2}) + (\ref{dd3}) - (\ref{dd4}) ~\Rightarrow \bar{D}_{23} + \bar{D}_{32} &=& M_2 + M_3 - M_1 \label{dd8} \\
(\ref{dd1}) + (\ref{dd3}) - (\ref{dd5}) ~\Rightarrow \bar{D}_{13} + \bar{D}_{31} &=& M_1 + M_3 - M_2  \label{dd9}
\end{eqnarray}
Combining with (\ref{dd1}) to (\ref{dd6}), we have the exact same form of the inequalities in (\ref{cd1}) to (\ref{cd9}). As $\bar{D}_{ji}\leq D_{ji}$,  (\ref{dd1}) to (\ref{dd6}) and (\ref{dd7}) to (\ref{dd9})  imply (\ref{cd1}) to (\ref{cd9}). This direction is proved. 

Next we consider the proof of Theorem \ref{theorem:new}.
We want to show that for a 3-user interference channel, if the rank of each interference link is symmetric, i.e., $D_{ji}=D_{ij}$, then Condition \eqref{1} (equivalently Condition \eqref{5}) is necessary for half-the-cake optimality. To prove this, it suffices to prove that when Condition \eqref{5} (inequalities (\ref{cd1}) to (\ref{cd9})) does not hold, we can always achieve more than half-the-cake DoF. We consider two cases, one when (\ref{cd7}) - (\ref{cd9}) is violated and the other when (\ref{cd1}) to (\ref{cd6}) is violated. We start with the first case.

\subsection{More than Half-the-cake when Inequalities (\ref{cd7}) - (\ref{cd9}) are violated}\label{subsec_case1}

As inequalities (\ref{cd7}) - (\ref{cd9}) are symmetric, without loss of generality, we assume (\ref{cd7}) is violated, i.e., 
\begin{eqnarray}
D_{12}+D_{21}<M_1+M_2-M_3 \label{vv}
\end{eqnarray}
We will show that 
%By a scheme combining zero-forcing, vector alignment and ergodic alignment, we can achieve 
$\tfrac{M_1+M_2+M_3+1}{2}$ DoF can be achieved, by generalizing the scheme of the counterexample in Appendix \ref{sec:example}.

\begin{figure}
%\begin{center}
\centering 
\includegraphics[scale =0.70]{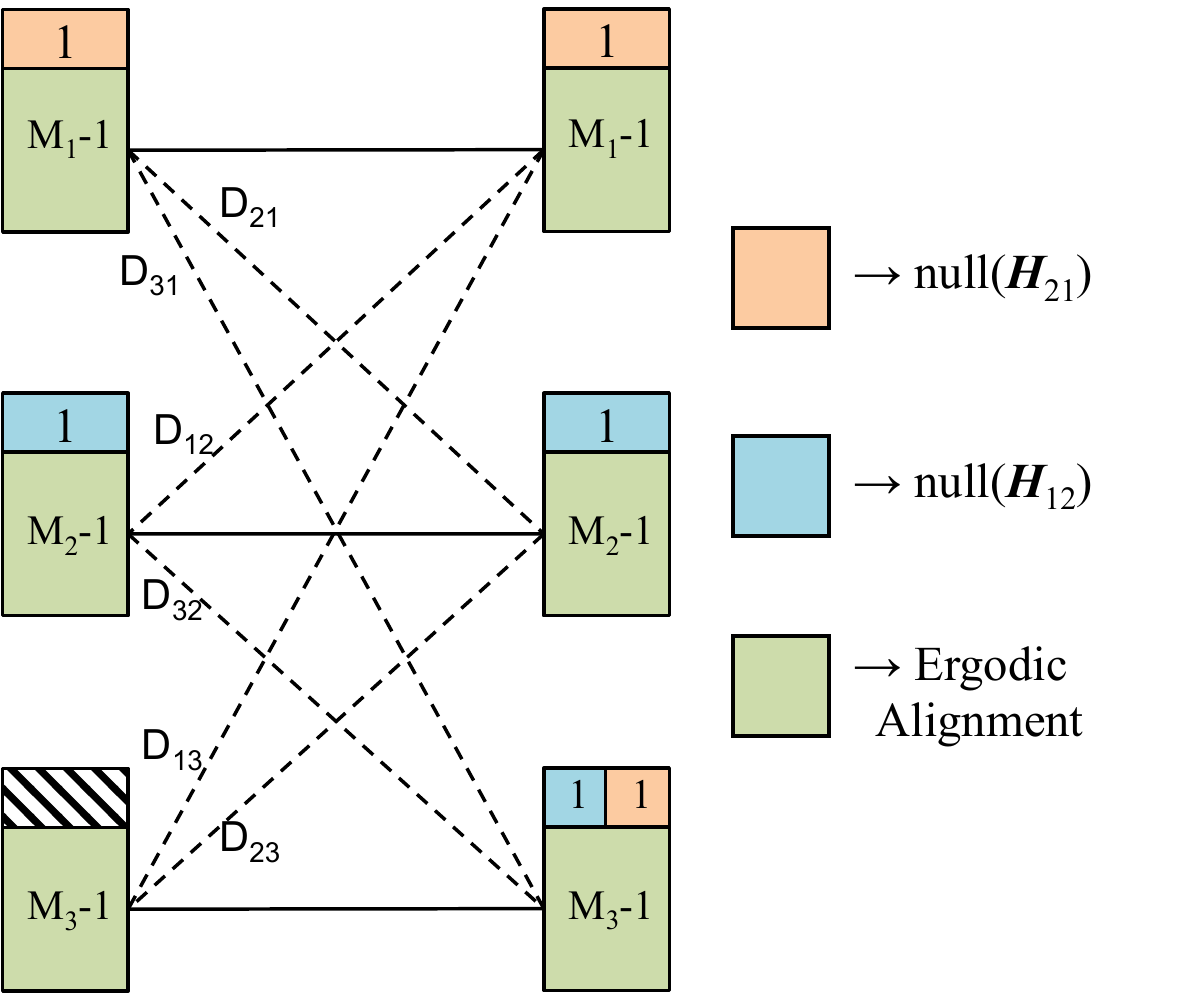}
\caption{Illustration of the scheme that achieves more than half-the-cake when $D_{12}+D_{21}<M_1+M_2-M_3$.}\label{example1}
%\end{center}
\end{figure}

The high level idea is the following. There exists a beamforming vector at Transmitter 1 and 2, respectively, that can align at Receiver 3 after being chosen from the null space they see at each other, as 
$M_1-D_{21}+M_2-D_{12}>M_3$ (refer to (\ref{vv})). 
So these two symbols occupy only 3 dimensions in total at all receivers (see Fig. \ref{example1} for an illustration). For the remaining $M_1+M_2+M_3-3$ dimensions, we apply ergodic alignment to achieve the DoF tuple $(\tfrac{M_1-1}{2}, \tfrac{M_2-1}{2}, \tfrac{M_3-1}{2})$  (green area in Fig. \ref{example1}). Added with the DoF tuple $(1, 1, 0)$ achieved as mentioned before, DoF tuple $(\tfrac{M_1+1}{2}, \tfrac{M_2+1}{2}, \tfrac{M_3-1}{2})$ is achieved in total. Thus, the sum-DoF value is more than half-the-cake.

Next we describe how to choose the beamforming vectors.
Specifically, we operate over 2 channel uses, where all cross channels remain the same, and all direct channels are generically different. We use $\mathbf{v}_{1}^{z}$ and $\mathbf{v}_{2}^{z}$ to denote the beamforming vectors of the signal at Transmitter 1 and Transmitter 2 that need to be aligned after zero-forcing. These signals are different over two channel uses. Mathematically, we have%the zero-forcing and alignment equations are as follows
\begin{align}
%\mathbf{H}_{21}\mathbf{v}_{1}^{z}&=\mathbf{0}\\
%\mathbf{H}_{12}\mathbf{v}_{2}^{z}&=\mathbf{0}\\
%\mathbf{H}_{31}\mathbf{v}_{1}^{z}&=\mathbf{H}_{32}\mathbf{v}_{2}^{z}
\mathbf{H}_{21}\mathbf{v}_{1}^{z} =\mathbf{0},
\mathbf{H}_{12}\mathbf{v}_{2}^{z} =\mathbf{0},
\mathbf{H}_{31}\mathbf{v}_{1}^{z} =\mathbf{H}_{32}\mathbf{v}_{2}^{z}
\end{align}
\begin{align}
\Rightarrow
\begin{matrix}\underbrace{\begin{bmatrix}
\mathbf{H}_{21} & \mathbf{0}      \\
\mathbf{0} & \mathbf{H}_{12} \\
\mathbf{H}_{31} & -\mathbf{H}_{32}
\end{bmatrix}_{(M_1+M_2+M_3)\times(M_1+M_2)}}\\ \mathbf{A}\end{matrix}
\begin{matrix}\underbrace{\begin{bmatrix}
\mathbf{v}_{1}^{z}   \\
\mathbf{v}_{2}^{z}
\end{bmatrix}}\\ \mathbf{v}\end{matrix}= \mathbf{0}. 
\end{align}

Note that matrix $\mathbf{A}$ is rank-deficient (sum of row ranks cannot be more than $D_{12}+D_{21}+M_3
%\min\{M_3, D_{31}+D_{32}\}
<M_1+M_2$, refer to (\ref{vv})), thus $\mathbf{v}$ can be determined as one basis vector of the right null space of $\mathbf{A}$. In the same manner, we can choose the received beamforming vectors $\mathbf{u}_{1}^{z}$ and $\mathbf{u}_{2}^{z}$ at Receiver 1 and Receiver 2 satisfying the following equations
\begin{align}
\mathbf{u}_{1}^{z}\mathbf{H}_{12}=\mathbf{0},
\mathbf{u}_{2}^{z}\mathbf{H}_{21}=\mathbf{0},
\mathbf{u}_{1}^{z}\mathbf{H}_{13}=\mathbf{u}_{2}^{z}\mathbf{H}_{23}.
\end{align}
%\begin{align*}
%\mathbf{U}_{1}^{z}\mathbf{H}_{12}&=\mathbf{0},\\
%\mathbf{U}_{2}^{z}\mathbf{H}_{21}&=\mathbf{0},\\
%\mathbf{U}_{1}^{z}\mathbf{H}_{13}&=\mathbf{U}_{2}^{z}\mathbf{H}_{23}.
%\end{align*}

Next, we use $\mathbf{V}_{k}^{e}$ and $\mathbf{U}_{k}^{e}$ to denote the $M_k\times (M_k-1)$ and $(M_k-1)\times M_k$ beamforming and filtering matrices at each transmitter and receiver, respectively. These matrices carry the signals for ergodic alignment, i.e., signals repeated by each user over two channel uses. User 1 and User 2 can choose $\mathbf{V}_{1}^{e}, \mathbf{V}_{2}^{e}$ and $\mathbf{U}_{1}^{e}, \mathbf{U}_{2}^{e}$ generically. %so that they are full rank and with no specific relation with the previously chosen vector. 
User 3 chooses its beamforming matrix as follows
\begin{eqnarray}
\mathbf{V}_{3}^{e}=\mathrm{span}(\mathrm{null}(\mathbf{u}_{2}^{z}\mathbf{H}_{23})), \mathbf{U}_{3}^{e}=\mathrm{span}(\mathrm{null}(\mathbf{H}_{32}\mathbf{v}_{2}^{z})).
\end{eqnarray}

As a result, %the beamforming vectors of each transmitter and receiver can be written as $\mathbf{V}_{3}=\mathbf{V}_{3}^{e}$, $\mathbf{U}_{3}=\mathbf{U}_{3}^{e}$, $\mathbf{V}_{k}=\begin{bmatrix}\mathbf{V}_{k}^{z} &\mathbf{V}_{k}^{e}\end{bmatrix}$, $\mathbf{U}_{k}=\begin{bmatrix}{\mathbf{U}_{k}^{z}}^T &{\mathbf{U}_{k}^{e}}^T\end{bmatrix}^T$, where $k=1, 2$. E
each receiver can eliminate interference by only subtracting the part of received signals corresponding to $\mathbf{U}_{k}^{e}$ over two channel uses. Thus, a total of $M_1+M_2+M_3+1$ DoF are achieved over two channel uses, which is more than half-the-cake DoF. The proof is complete. Remarkably, note that this proof does not require the assumption of symmetry, $D_{ij}\neq D_{ji}$, so it works for asymmetric settings as well.

%\subsection{More than Half-the-cake when Inequalities (\ref{cd7}) - (\ref{cd9}) are violated}\label{subsec_case1}
\subsection{More than Half-the-cake when Inequalities (\ref{cd1}) - (\ref{cd6}) are violated}\label{subsec_case2}
We now consider the case where  (\ref{cd1}) - (\ref{cd6}) are violated. Without loss of generality, we assume (\ref{cd1}) is violated, i.e., 
\begin{eqnarray}
D_{12}+D_{13}<M_1.
\end{eqnarray}
Since the ranks of the interference channels are symmetric, we have $D_{12}=D_{21}$ and $D_{13}=D_{31}$. Thus
\begin{eqnarray}
D_{21}+D_{31}<M_1.\label{vvv}
\end{eqnarray}
We will show that 
$\tfrac{M_1+M_2+M_3+1}{2}$ DoF can be achieved, by combining zero-forcing and ergodic alignment. 

This case turns out to be quite simple.
The high level idea is the following. There exists a beamforming vector at Transmitter 1 that cannot be seen by both Receivers 2 and 3. The symbol carried by this vector occupies only 1 dimension in total at all receivers. For the remaining $M_1+M_2+M_3-1$ dimensions, we apply ergodic alignment to achieve the DoF tuple $(\tfrac{M_1-1}{2}, \tfrac{M_2}{2}, \tfrac{M_3}{2})$. Added with the DoF tuple $(1, 0, 0)$ achieved as mentioned above, DoF tuple $(\tfrac{M_1+1}{2}, \tfrac{M_2}{2}, \tfrac{M_3}{2})$ is achieved in total. Thus, the sum-DoF value is more than half-the-cake.

Next we proceed to describe the scheme.
Specifically, we operate over 2 channel uses, where all cross channels remain the same, and all direct channels are generically different. We use $\mathbf{v}_{1}^{z}$ to denote the beamforming vector of the signal at Transmitter 1 that is zero-forced at Receivers 2 and 3. This signal is different over two channel uses. Mathematically, we have%the zero-forcing and alignment equations are as follows
\begin{align}
%\mathbf{H}_{21}\mathbf{v}_{1}^{z}&=\mathbf{0}\\
%\mathbf{H}_{12}\mathbf{v}_{2}^{z}&=\mathbf{0}\\
%\mathbf{H}_{31}\mathbf{v}_{1}^{z}&=\mathbf{H}_{32}\mathbf{v}_{2}^{z}
\begin{bmatrix}
\mathbf{H}_{21}&\mathbf{H}_{31}\end{bmatrix}\mathbf{v}_{1}^{z} ={0}.
\end{align}
%\begin{align}
%\Rightarrow
%\begin{matrix}\underbrace{\begin{bmatrix}
%\mathbf{H}_{21} & \mathbf{0}      \\
%\mathbf{0} & \mathbf{H}_{12} \\
%\mathbf{H}_{31} & -\mathbf{H}_{32}
%\end{bmatrix}_{(M_1+M_2+M_3)\times(M_1+M_2)}}\\ \mathbf{A}\end{matrix}
%\begin{matrix}\underbrace{\begin{bmatrix}
%\mathbf{v}_{1}^{z}   \\
%\mathbf{v}_{2}^{z}
%\end{bmatrix}}\\ \mathbf{v}\end{matrix}= \mathbf{0}. 
%\end{align}

Note that matrix $\begin{bmatrix}\mathbf{H}_{21}&\mathbf{H}_{31}\end{bmatrix}$ is rank-deficient (the rank cannot be more than $D_{21}+D_{31}<M_1$, refer to (\ref{vvv})), thus $\mathbf{v}_{1}^{z}$ can be determined as one basis vector of the right null space of $\begin{bmatrix}\mathbf{H}_{21}&\mathbf{H}_{31}\end{bmatrix}$. 
In the same manner, we can choose the received beamforming vectors $\mathbf{u}_{1}^{z}$ at Receiver 1 such that %satisfying %the following equation
%\begin{align}
$\mathbf{u}_{1}^{z}\begin{bmatrix}\mathbf{H}_{12}&\mathbf{H}_{13}\end{bmatrix} = {0}.$
%\mathbf{u}_{1}^{z}\mathbf{H}_{12}=\mathbf{0},
%\mathbf{u}_{2}^{z}\mathbf{H}_{21}=\mathbf{0},
%\mathbf{u}_{1}^{z}\mathbf{H}_{13}=\mathbf{u}_{2}^{z}\mathbf{H}_{23}.
%\end{align}
%\begin{align*}
%\mathbf{U}_{1}^{z}\mathbf{H}_{12}&=\mathbf{0},\\
%\mathbf{U}_{2}^{z}\mathbf{H}_{21}&=\mathbf{0},\\
%\mathbf{U}_{1}^{z}\mathbf{H}_{13}&=\mathbf{U}_{2}^{z}\mathbf{H}_{23}.
%\end{align*}

Next, we use $\mathbf{V}_{1}^{e}$ and $\mathbf{U}_{1}^{e}$ to denote the $M_k\times (M_k-1)$ and $(M_k-1)\times M_k$ beamforming and filtering matrices at Transmitter 1 and Receiver 1, respectively. For $k \in\{1, 2\}$, we use $\mathbf{V}_{k}^{e}$ and $\mathbf{U}_{k}^{e}$ to denote the $M_k\times M_k$ beamforming and filtering matrices at Transmitter $k$ and Receiver $k$, respectively. These matrices carry the signals for ergodic alignment, i.e., signals repeated by each user over two channel uses. Each user can choose its beamforming and filtering matrices generically. 

As a result, %the beamforming vectors of each transmitter and receiver can be written as $\mathbf{V}_{3}=\mathbf{V}_{3}^{e}$, $\mathbf{U}_{3}=\mathbf{U}_{3}^{e}$, $\mathbf{V}_{k}=\begin{bmatrix}\mathbf{V}_{k}^{z} &\mathbf{V}_{k}^{e}\end{bmatrix}$, $\mathbf{U}_{k}=\begin{bmatrix}{\mathbf{U}_{k}^{z}}^T &{\mathbf{U}_{k}^{e}}^T\end{bmatrix}^T$, where $k=1, 2$. E
each receiver can eliminate interference by only subtracting the part of received signals corresponding to $\mathbf{U}_{k}^{e}$ over two channel uses. Thus, a total of $M_1+M_2+M_3+1$ DoF are achieved over two channel uses, which is more than half-the-cake DoF. The proof is complete.

\section{Non-necessity of Condition \eqref{1} at ``Boundary Cases''}%Half-the-Cake is Still Optimal if Optimality Condition is Violated} 
\label{proof:last}

\subsection{Proof of Theorem \ref{special2}: $M_1=M_2+M_3$}\label{subsec_specialcase2}
Consider a 3-user interference channel where $M_1=M_2+M_3$, We want to show that if $D_{12} = M_2, D_{13} = M_3$ or $D_{21} = M_2, D_{31} = M_3$, then $d_\Sigma = \frac{1}{2} M_\Sigma$. Achievability is implied by Theorem \ref{theorem:ach}, so we proceed to the outer bound.
Since we are considering the outer bound, cooperation between the users will not hurt. Therefore, we allow Transmitter 2 and Transmitter 3 to cooperate and they form a new Transmitter $2'$. Similarly, we allow Receiver 2 and Receiver 3 to cooperate and they form a new Receiver $2'$. We now arrive at a 2-user interference channel, where Transmitter/Receiver 1 has $M_1$ antennas and Transmitter/Receiver $2'$ has $M_{2'} = M_2 + M_3$ antennas. The desired channels have full rank, the interference channel from Transmitter 1 to Receiver $2'$ has rank $D_{2'1} = D_{21} + D_{31}$, and  the interference channel from Transmitter $2'$ to Receiver $1$ has rank $D_{12'} = D_{12} + D_{13}$. For such a rank-deficient 2-user MIMO interference channel, we invoke Theorem 1 in \cite{krishnamurthy2015degrees} to obtain the following outer bound which also serves as outer bound for the original 3-user interference channel, $d_\Sigma \leq M_1 + M_{2'} - \max(D_{2'1}, D_{12'}) = M_1 + M_2 + M_3 - \max(D_{21} + D_{31}, D_{12} + D_{13})$. Therefore, if $D_{12} = M_2, D_{13} = M_3$ or $D_{21} = M_2, D_{31} = M_3$, the outer bound becomes $d_\Sigma \leq M_1 = \frac{1}{2}M_\Sigma$. This completes the proof.

\subsection{Proof of Theorem \ref{general3user}: $M_1=M_2$}\label{subsec_specialcase1}

\begin{figure}[h]
\centering
\subfigure[]{
\label{3user1}
\includegraphics[width=0.28\textwidth]{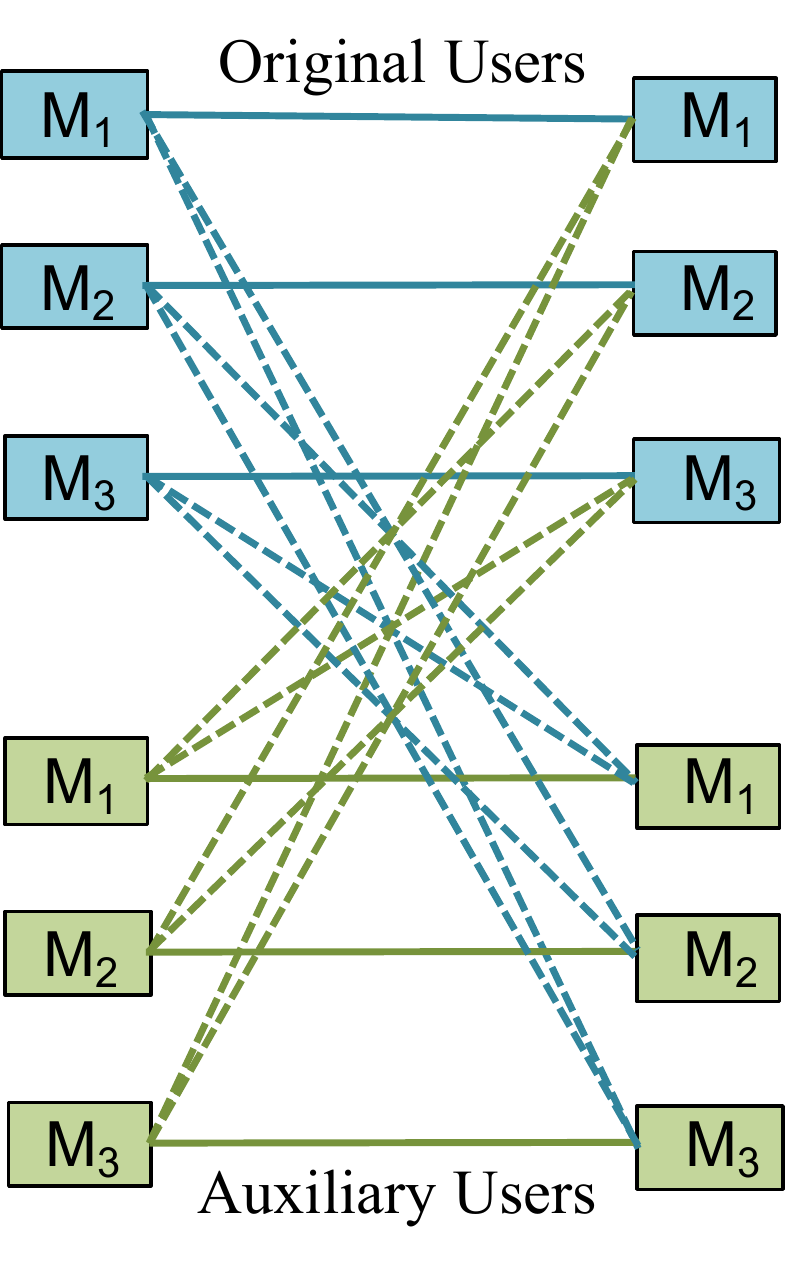}}
\hspace{1.2in}
\subfigure[]{
\label{3user2}
\includegraphics[width=0.32\textwidth]{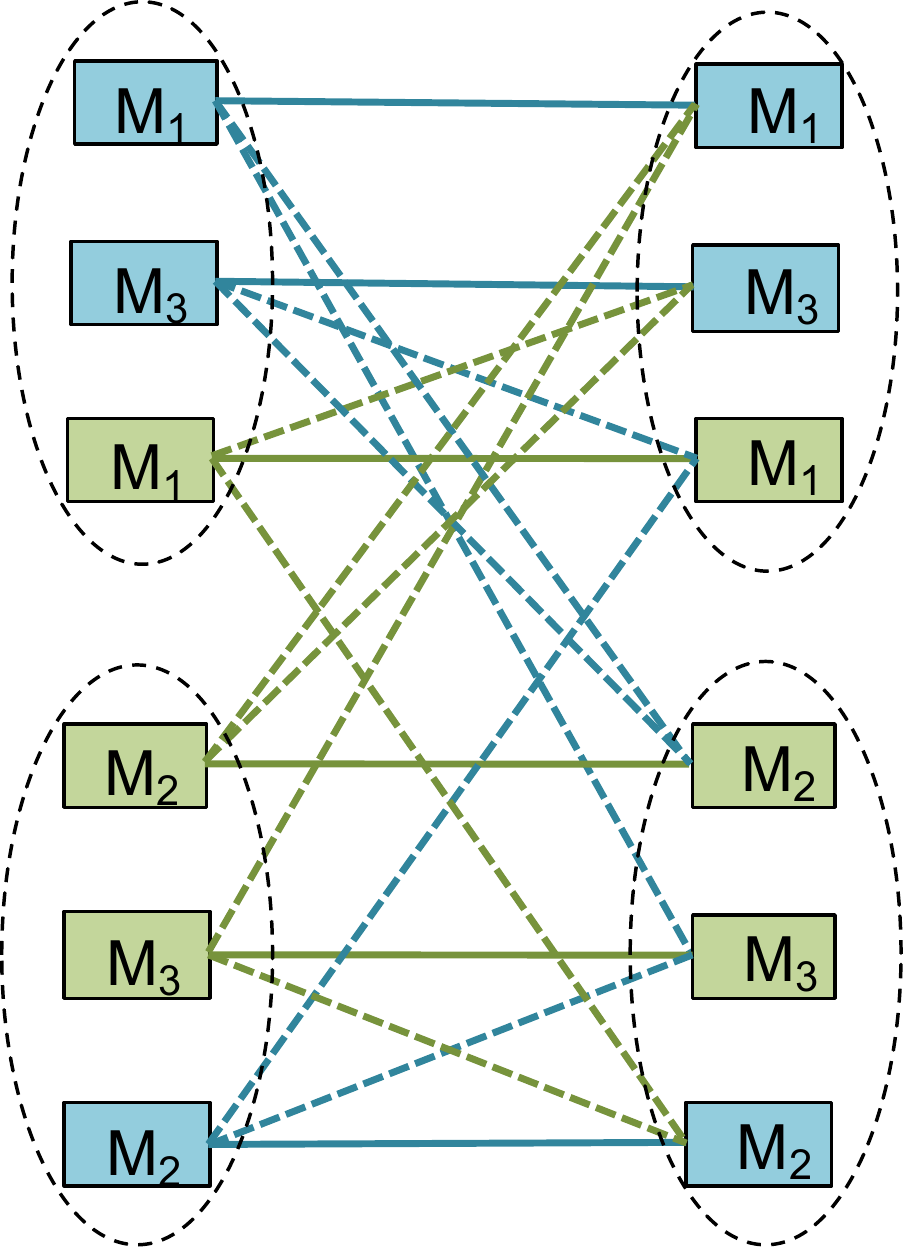}}
\caption[]{
\subref{3user1} A $6$-user interference channel created by adding an auxiliary user for each user in the original 3-user channel, and \subref{3user2} Illustration of users' cooperation in this new channel.}
\label{Fig.3user}
\end{figure}

Consider a 3-user interference channel where $M_1=M_2$. We want to show that if $D_{21}=M_{1}, D_{31}=D_{23}=M_{3}$ or $D_{12}=M_{1}, D_{13}=D_{32}=M_{3}$, then $d_\Sigma = \frac{1}{2} M_\Sigma$. Achievability is implied by Theorem \ref{theorem:ach}, so we proceed to the outer bound. For such a purpose, we create a 6-user interference channel by adding an auxiliary User $k'$ for each Original User $k$. %, and choosing each new channel in the same manner as in Theorem \ref{theorem:LDoF}.  
We denote the channels in the new network by notations with hat symbol, e.g., $\hat{\bf H}_{ji'}$, and the channels in the original network by notations with no hat symbol, e.g., ${\bf H}_{ji}$. The channels in the new network are chosen in the same manner as in Theorem \ref{theorem:LDoF}, i.e., $ \forall i,j \in \{1,2,3\}$,  1) $\hat{\bf H}_{j'i}=\hat{\bf H}_{ji'}= {\bf H}_{ji}$ whenever $i\neq j$, 2) $ \hat{\bf H}_{i'i'}=\hat {\bf H}_{ii}= {\bf H}_{ii}$, 3) $ \hat{\bf H}_{j'i'}=\hat{\bf H}_{ji}$ is the matrix of zeros whenever $i\neq j$, and 4) $\hat{\bf H}_{i'i}=\hat{\bf H}_{ii'}$ is the matrix of zeros.  See Fig. \ref{Fig.3user} for a pictorial illustration.
By this construction, any coding scheme for the original channel still works if each auxiliary User $i'$ uses the same codebook as User $i$. Therefore the sum-DoF value of this new network is at least twice that of the original network. Now in this new network, we allow User 1, User 3 and User $1^\prime$ to cooperate, and User $2^\prime$, User $3^\prime$ and User 2 to cooperate, which can only help. This creates a 2-user interference channel where the first transmitter/receiver has $2M_1+M_3$ antennas, the second transmitter/receiver has $2M_2+M_3$ antennas. We denote the interference channel between the first/second transmitter and the second/first receiver as ${\bf \bar{H}}_{21}$ and ${\bf \bar{H}}_{12}$, respectively. Note that as $M_1 = M_2$, both ${\bf \bar{H}}_{21}$ and ${\bf \bar{H}}_{12}$ are square matrices. They may be written as
\begin{equation}\label{eq:3user}
{\bf \bar{H}}_{21}=\bordermatrix{&\mbox{\tiny $M_1$}&\mbox{\tiny $M_3$}&\mbox{\tiny $M_1$}\cr
\mbox{\tiny $M_2$}&{\bf H}_{21}&{\bf H}_{23}&{\bf 0}\cr
 \mbox{\tiny $M_3$}&{\bf H}_{31}&{\bf 0}&{\bf 0}\cr
 \mbox{\tiny $M_2$}&{\bf 0}&{\bf 0}&{\bf H}_{21}\cr
 }
 \end{equation}
 \begin{equation}\label{eq:3user2}
{\bf \bar{H}}_{12}=\bordermatrix{&\mbox{\tiny $M_2$}&\mbox{\tiny $M_3$}&\mbox{\tiny $M_2$}\cr
\mbox{\tiny $M_1$}&{\bf H}_{12}&{\bf H}_{13}&{\bf 0}\cr
 \mbox{\tiny $M_3$}&{\bf H}_{32}&{\bf 0}&{\bf 0}\cr
 \mbox{\tiny $M_1$}&{\bf 0}&{\bf 0}&{\bf H}_{12}\cr
 }
 \end{equation}
If ${\bf \bar{H}}_{21}$ has full rank, then the first receiver, after decoding its desired signal, can subtract it out and then proceed to decode the interfering signal as well (subject to noise distortion, inconsequential for DoF). Thus, the sum-DoF of the interference channel cannot be more than $2M_1+M_3=M_\Sigma$, and therefore the sum-DoF of the original network cannot be more than $\frac{1}{2}M_\Sigma$. Similarly, if ${\bf \bar{H}}_{12}$ has full rank, then the second receiver can decode both messages such that $d_\Sigma \leq \frac{1}{2}(2M_2+M_3) = \frac{1}{2}M_\Sigma$.
We are left to prove that if $D_{21}=M_{1}, D_{31}=D_{23}=M_{3}$, then  ${\bf \bar{H}}_{21}$ has full rank and symmetrically, if $D_{12}=M_{1}, D_{13}=D_{32}=M_{3}$, then ${\bf \bar{H}}_{12}$ has full rank. We prove the first statement and the second follows similarly. We prove that when $D_{21}=M_{1}, D_{31}=D_{23}=M_{3}$, the determinant polynomial of ${\bf \bar{H}}_{21}$ is not identically zero. It suffices to find one channel realization such that the determinant polynomial is not zero. The channels we construct are as follows.
\begin{align*}
{\bf H}_{21}&%={\bf H}_{12}
={\bf I}_{M_1},\\
{\bf H}_{31}&%={\bf H}_{32}
=\begin{bmatrix}
{\bf I}_{M_3}& {\bf 0}_{M_3\times(M_1-M_3)}
\end{bmatrix},\\
{\bf H}_{23}&%={\bf H}_{13}
=\begin{bmatrix}
{\bf I}_{M_3}\\
{\bf 0}_{(M_2-M_3)\times M_3}
\end{bmatrix}. \\
\end{align*}
Note that the rank constraints are satisfied and it is easily seen that the determinant of  ${\bf \bar{H}}_{21}$ is non-zero. Therefore,  ${\bf \bar{H}}_{21}$ has full rank almost surely. We now finish the proof of the outer bound. Note that the procedure is a specific realization of Theorem \ref{theorem:general_k}.
Combined with the achievability, the proof is complete.

%\bibliographystyle{IEEEtran}
%\bibliography{IEEEabrv,myrefbib}
\bibliography{myrefbib}

\end{document}